\documentclass[aps,prb,amsmath,twocolumn,floatfix]{revtex4}

\usepackage{epsf}
\usepackage{array}

\def\e{\varepsilon}

\def\r{{\bf r}}

\def\n{{\bf n}}
\def\p{{\vec p}}
\def\ps{{\vec p\ '}}
\def\q{{\bf q}}
\def\o{\omega}

\def\gs{{\gamma_{\rm s}}}
\def\ts{{\tau_{\rm s}}}
\def\gso{{\gamma_{\rm so}}}
\def\fp{\tau_{\rm e}}
\def\tK{\tau}
\def\os{\omega_{S}}
\def\Dt{\Delta t}
\def\Df{\Delta \Phi}
\def\f0{\Phi_0}
\def\Tm{T_{\rm m}}
\def\gd{\gamma_{\varphi}}
\def\muB{\mu_{\rm B}}
\def\gH{\gamma_{\rm orb}}

\def\Sodin{\langle S_z\rangle}
\def\S2{\langle S_z^2\rangle}
\begin{document}
\title{Conductance of Mesoscopic Systems with Magnetic Impurities}
\author{M.G. Vavilov and L.I. Glazman }
\affiliation{Theoretical Physics Institute, University of Minnesota,
Minneapolis, MN 55455}
\date{\today}

\begin{abstract}  
We investigate the combined effects of magnetic impurities and
applied magnetic field on the interference contribution to
the conductance of disordered metals. We show that in a metal with weak
spin-orbit interaction, the polarization of impurity spins reduces
the rate of electron phase relaxation, thus enhancing the weak
localization correction to conductivity.  Magnetic field also
suppresses thermal fluctuations of magnetic impurities, leading
to a recovery of the conductance fluctuations. This recovery occurs
regardless the strength of the spin-orbit interaction. We calculate the
magnitudes of the weak localization correction and of the mesoscopic
conductance fluctuations at an arbitrary level of the spin
polarization induced by a magnetic field. Our analytical results for
the ``$h/e$'' Aharonov-Bohm conductance oscillations in metal rings can
be used to extract spin and gyromagnetic factor of  magnetic impurities 
from existing experimental data.
\end{abstract}
\maketitle

%\begin{multicols}{2}
%\narrowtext
%%%%%%%%%%%%%%%%%%%%%%%%%%%%%%%%%%%%%%%%%%%%%%%%%%%%%%%%%%%%%%%%%%%%%%%%%%%%%%
%%%%%%%%%%%%%%%%%%%%%%%%%%%%%%%%%%%%%%%%%%%%%%%%%%%%%%%%%%%%%%%%%%%%%%%%%%%%%%
%%%%%%%%%%%%%%%%%%%%%%%%%%%%%%%%%%%%%%%%%%%%%%%%%%%%%%%%%%%%%%%%%%%%%%%%%%%%%%
\section{Introduction}
\label{sec:1}
%%%%%%%%%%%%%%%%%%%%%%%%%%%%%%%%%%%%%%%%%%%%%%%%%%%%%%%%%%%%%%%%%%%%%%%%%%%%%%
%%%%%%%%%%%%%%%%%%%%%%%%%%%%%%%%%%%%%%%%%%%%%%%%%%%%%%%%%%%%%%%%%%%%%%%%%%%%%%
%%%%%%%%%%%%%%%%%%%%%%%%%%%%%%%%%%%%%%%%%%%%%%%%%%%%%%%%%%%%%%%%%%%%%%%%%%%%%%

Conductance of disordered metals is sensitive to the applied magnetic
field. At weak magnetic field the conductivity of a bulk metal has a sharp
peak due to the weak localization (WL).\cite{WL}
%\cite{GLK,AAKL,experiment1} 
Similarly, the conductance
of mesoscopic metals fluctuates as the magnetic field is
changing.\cite{A1,experiment2} Both the conductance fluctuations and
the WL correction to the conductivity are quantum
mechanical phenomena originating from the interference of quantum
states. As any other interference phenomena, they may be 
demolished by interaction processes.\cite{AAK}

Localized spins affect the electron transport in metals.  Various
properties of electron kinetics are sensitive to different aspects of
the localized spin dynamics. The energy exchange between electrons in
the process of scattering off a magnetic impurity is made possible by
the quantum fluctuations of the impurity spin: its virtual flip in the
course of scattering facilitates the energy transfer between the two
electrons.\cite{KG}  On the other hand, no spin dynamics of
impurities is needed for the suppression of the weak localization
correction to conductivity; interaction of electron spins with randomly oriented magnetic moments already leads to that
suppression.\cite{HLN}

Universal conductance fluctuations (UCF) are
not suppressed by static magnetic moments.  However, even a relatively
slow relaxation  of
individual magnetic moments, such as Korringa relaxation,\cite{Korringa} leads to the time-averaging of the random
potential ``seen'' by transport electrons in the course of
measurement, and the mesoscopic fluctuations of the dc conductance get
averaged out.\cite{BFK,Hershfield} The sensitivity of conductance fluctuations to the time
evolution of the system of localized magnetic
moments~\cite{Altshuler3} was used extensively to probe the spin-glass
freezing in metals~\cite{vegvar,cumn} and semiconductors.\cite{jarozunski}

An applied magnetic field may be used to control
the statistical properties and dynamics of localized magnetic
moments. It was noticed~\cite{cumn} that in
a strong magnetic field the amplitude of conductance fluctuations and ``$h/e$'' Aharonov-Bohm
(AB) oscillations increases, apparently because the spin fluctuations
are quenched.\cite{BFK} Recently the dependence of the amplitude of
``$h/e$'' AB oscillations on magnetic field was carefully measured on Cu
wire rings.\cite{Birge} The goal of that measurement was to
corroborate the existence of localized spins, conjectured on the basis
of measurements of the electron energy relaxation rate in Cu wires.\cite{PothierEtal} 

Theory of conductance fluctuations  and WL correction to the conductivity at partial spin polarization of magnetic impurities
has not been developed  yet. 
Only the limits of no spin polarization at $B=0$ and of strong
polarization at $B\gg T/g\muB S$ were considered\cite{BFK,Amaral,F-cf,meyer}  (here $B$ is the magnetic
field, $S$ is the impurity spin, $T$ is the system temperature, $g$ is the impurity gyromagnetic factor, and  $\muB$ is the Bohr magneton).

In this paper we concentrate on the interference contribution to the
linear conductance of mesoscopic systems in the presence of partially polarized magnetic impurities. Particularly, we calculate the weak localization correction to the conductivity and the amplitude of the mesoscopic conductance fluctuations.

%%%%%%%%%%%%%%%%%%%%%%%%%%%%%%%%%%%%%%%%%%%%%%%%%%%%%%%%%%%%%%%%
%%%%%%%%%%%%%%%%%%%%%%%%%%%%%%%%%%%%%%%%%%%%%%%%%%%%%%%%%%%%%%%%
\section{Main results}
\label{sec:2}
%%%%%%%%%%%%%%%%%%%%%%%%%%%%%%%%%%%%%%%%%%%%%%%%%%%%%%%%%%%%%%%%
%%%%%%%%%%%%%%%%%%%%%%%%%%%%%%%%%%%%%%%%%%%%%%%%%%%%%%%%%%%%%%%%
It is well known that scattering of electrons off magnetic impurities in
the absence of a magnetic field suppresses the interference correction to
the conductivity of a wire:\cite{WL,HLN}
\begin{equation}
\Delta\sigma=-\frac{e^2}{2\pi\hbar }
\sqrt{D}
\left(
\frac{3}{\sqrt{2/3\tau_{\rm s}}}-
\frac{1}{\sqrt{2/\tau_{\rm s}}}
\right).
\label{eq:1.0}
\end{equation}
Here $D$ is the diffusion constant for electrons in the wire and  $1/\ts$ is the electron scattering rate off magnetic impurities.
To the lowest order in the exchange constant $J$
calculation, $1/\ts=2\pi\nu n_{\rm s}J^2S(S+1)$, where $n_{\rm s}$ is the
concentration of magnetic impurities, $\nu$ is the electron density of
states at the Fermi level per spin degree of freedom, and $S$ is the magnitude of impurity spins. Taking into
account the Kondo renormalization of the exchange constant, at
temperatures well above the Kondo temperature $T_K$, this rate can be cast
in the form
\begin{equation}
\frac{1}{\ts}=\frac{8\pi n_{\rm s}}{\nu}\frac{S(S+1)}{\ln^2 T/T_K},
\quad T\gg T_K. 
\label{eq:1.1}
\end{equation}
It is clear from Eq.~(\ref{eq:1.1}) that the phase relaxation rate
increases upon reduction of the temperature towards $T_K$. This rate
reaches maximum at $T\sim T_K$, and decreases if the temperature is further
reduced. The specific form of function $\ts(T)$ at low temperatures depends on the spin $S$
of local moments. The screening of the local spin $S=1/2$ is complete at
$T=0$ and at $T\ll T_K$ the  phase relaxation rate  can be found from
the Nozi\'eres' Fermi liquid theory,
\begin{subequations}
\label{eq:1.2+3}
\begin{equation}
\frac{1}{\ts}\propto \frac{n_{\rm s}}{\nu}\left(\frac{T}{T_K}\right)^2,
\quad T\ll T_K.
\label{eq:1.2}
\end{equation}
If $S\geq 1$, then screening is incomplete, and the logarithmic
renormalization theory can be employed for the evaluation of $\ts$ at
low temperatures,\cite{finiteS}
\begin{equation}
\frac{1}{\ts}=\frac{8\pi n_{\rm s}}{\nu}\frac{S^2-1/4}{\ln^2 T_K/T},
\quad T\ll T_K. 
\label{eq:1.3}
\end{equation}
\end{subequations}
Note, that the low-temperature asymptotes, Eqs.~(\ref{eq:1.2+3}), differ from the approximate expression, used sometimes in the analysis of experiments.\cite{NSuhl,NSuhl1}

In this section, we only present the result for WL correction $\Delta
\sigma$ to the conductivity of a thin wire in the absence of spin-orbit
scattering. If, in addition, the density of magnetic impurities is small,
then starting from rather weak fields the condition $g\mu_BB\tau_s\gg 1$
is satisfied, and we find
\begin{equation}
\Delta\sigma  = 
-\frac{e^2}{4\pi\hbar }
\int\frac{d\e/T}{\cosh^2\e/2T}
\frac{\sqrt{D}}{\sqrt{\Gamma(\e) + \vartheta D A B^2/\Phi_0^2}}.
\label{eq:1.4}
\end{equation} 
Here $A$ is the wire cross-section area, $\vartheta$ is a numerical
factor, and $\Phi_0=hc/2e$. Function $\Gamma(\e)$ represents the spin flip rate in the presence of magnetic field $B$:
\begin{equation}
\Gamma(\e)=\left[1-\frac{\langle \hat S_z^2 \rangle + 
                    \langle \hat S_z\rangle\tanh(\e+g\mu_BB)/2T}
                    {S(S+1)}\right]\frac{1}{\ts},
\label{eq:1.5}
\end{equation}
Here $\langle\dots\rangle$ stands for the thermodynamic average over the
states of an isolated impurity spin, see Eq.~(\ref{eq:5_5a}). The 
Kondo-renormalized rate $1/\tau_{\rm s}$ in Eq.~(\ref{eq:1.5}) is given by
Eq.~(\ref{eq:1.1}) at weak magnetic field, $B<T/g\muB$, and by
\begin{equation}
\frac{1}{\ts}=\frac{2\pi n_{\rm s}}{\nu}\frac{S(S+1)}
{\ln^2[g\muB B/T_K]}
\label{eq:1.1a}
\end{equation}
at strong magnetic field, $B>{\rm max}\{T,T_K\}/g\muB$.  

%At $B\ll T/g\mu_B$ spins are not polarized, and
%$\Gamma(\varepsilon)=2/3\ts$. 

In the domain of relatively weak magnetic fields, 
$1/g\mu_B\tau_s\ll B \ll T/g\muB$, the spins are not polarized, however, their precession already affects the value of WL correction. Equation~(\ref{eq:1.4}), valid in this domain, yields $\Delta\sigma$ differing from the full zero-field value (\ref{eq:1.0}) by a factor $2/(3-\sqrt{1/3})\approx 0.83$. We discuss the detailed
behavior of the WL correction in the field region $B\lesssim 1/g\muB\ts$
in Section IV. The main variation of the WL correction, however, is associated with the spin polarization. This variation occurs at $B\sim T/g\muB$, and is described by Eqs.~(\ref{eq:1.4}) and (\ref{eq:1.5}).

At strong magnetic field, $B\gtrsim
T/g\muB$, the spin flip rate of thermal electrons ($\varepsilon\sim T$)
becomes exponentially small, 
$\Gamma\propto \exp(-g\muB B/T)/[(S+1)\tau_{\rm_s}]$.  
In this case higher-order terms in the exchange
interaction must be retained to calculate the phase relaxation rate. 
We find the following result for $\Gamma(\varepsilon)$:
\begin{equation}
\Gamma(\varepsilon)=\frac{1}{2\pi}\frac{\nu}{n_{\rm s}\tau_{\rm s}^2}
\frac{\pi^2T^2+\varepsilon^2}{[g\mu_BB(S+1)]^2}.
\label{eq:1.6}
\end{equation}
The crossover between the rates given by Eqs.~(\ref{eq:1.5}) and
(\ref{eq:1.6}) occurs at 
\begin{equation}
B_*(T) \approx (T/g\mu_B)\ln (\nu/Sn_{\rm s}\tau_{\rm s}).
\label{eq:1.7} 
\end{equation}
The relaxation rate at high fields, Eq.~(\ref{eq:1.6}),
may still be large enough to compete with the conventional
mechanisms\cite{AAK,WL} of phase relaxation caused by the electron-electron and electron-phonon interactions.

If one disregards the orbital effect of magnetic field (very thin wire)
then the WL correction to the conductivity, Eq.~(\ref{eq:1.4}), is a function of the ratio $B/T$, up to slowly varying logarithmic factors $\ln(g\mu_BB/T_K)$ and $\ln(T/T_K)$, see Eqs.~(\ref{eq:1.1}) and (\ref{eq:1.1a}).   
%A detailed analysis of the case $S=1/2$ can be
%found in Sec.~\ref{sec:4.3}, see Eqs.~(\ref{eq:wl1:2f})--(\ref{eq:cse1:2})
%and (\ref{eq:wl1:2imp}).

\begin{figure}
\centerline{\epsfxsize=8cm
\epsfbox{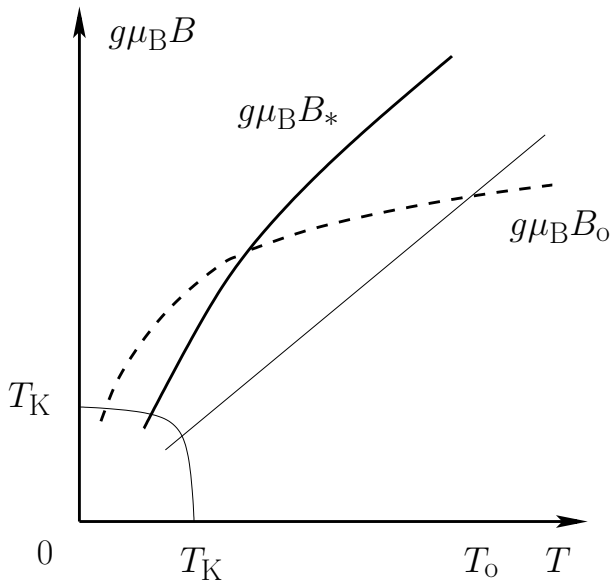}}
\caption {
Different regimes of the temperature, $T$, and magnetic field, $B$, dependence of the mesoscopic conductance fluctuations and the WL correction to conductivity.
Below the Kondo temperature, $T{\protect\lesssim} T_K$, and at weak magnetic field, $g\mu_B B{\protect\lesssim} T_K$, magnetic impurities are screened and the electron phase relaxation rate is given by Eqs.~(\protect\ref{eq:1.2+3}). The impurity magnetic moments acquire a significant polarization at fields around $B= T/g\muB S$ (straight solid line), which results in strong dependence of  the electron phase relaxation rate on the applied magnetic field. The exponential behavior of the phase relaxation rate on $B/T$ is replaced by a weaker power-law function, see Eq.~(\ref{eq:1.6}), at a crossover field $B_*$ defined by Eq.~(\ref{eq:1.7}) and depicted by the solid bold curve. Strong variation of the WL correction is possible only in the absence of spin-orbit interaction and at sufficiently weak orbital effect of the magnetic field. This last condition is satisfied below the dashed curve
$B_{\rm o}(T)$ defined by Eq.~(\ref{eq:19}), which intersects the line 
$g\muB SB=T$ at temperature $T_{\rm o}$.}
\label{fig:TB-plane}
\end{figure}

Different regimes for the WL are
schematically represented in Fig.~1. Substantial variation of $\Delta\sigma$
with $B$ occurs at $B\gtrsim T/g\muB S$. The curve $B=B_*(T)$ is shown in Fig.~\ref{fig:TB-plane} by the bold solid 
line.

The effect of magnetic-field-induced polarization of localized spins on
the WL correction to the conductivity is significant only if two quite stringent conditions are met. 

First, the orbital effect of the magnetic field $B$ on
WL~\cite{BFK} is needed to be small in the range of fields $B\gtrsim
T/g\muB$, where the spin polarization is significant.
The orbital effect dominates over the spin scattering, if $DA B^2/\Phi^2_0\lesssim\Gamma (\varepsilon\simeq T)$, see Eq.~(\ref{eq:1.4}). This condition defines the upper bound (a dashed line in Fig.~\ref{fig:TB-plane}) for the domain
of fields $B$ in which the spin polarization noticeably affects the WL
correction. In general, to avoid the undesirable orbital effect, measurements are to be done on thin wires at low temperature.

Second,  the spin-orbit
scattering must be sufficiently weak. This condition limits the range of host conductors to materials with sufficiently low atomic numbers.\cite{WL} 
In heavier materials, the spin-orbit interaction suppresses the triplet part of the WL correction to the conductivity, represented by the first term in Eq.~(\ref{eq:1.0}). Polarization of local moments does not eliminate the
randomness of the effective magnetic field induced by spin-orbit
interaction, and the triplet contribution to the WL remains suppressed. At the same time, the singlet part of the weak (anti)localization correction [the second term in Eq.~(\ref{eq:1.0})] is insensitive to the spin polarization.

The effect of spin polarization on the mesoscopic conductance fluctuations is not subject to the two restrictions discussed above. In other words, the manifestation of the spin polarization in the conductance fluctuations and in the ``$h/e$'' Aharonov-Bohm effect is much more robust than that in the WL. 
The presence of magnetic impurities suppresses the conductance
fluctuations only due to the spin dynamics of the impurity system. If the measurement time is significantly longer than the characteristic time
of the variation of impurity spin configuration, then the conductance
fluctuations are averaged out. Quenching of the spin configuration by the applied magnetic field decreases  the effect of the conductance
averaging and restores the fluctuations.
In the limit of full polarization, magnetic
impurities no longer affect the conductance fluctuations.\cite{BFK} 

We concentrate on
the amplitude of the Aharonov-Bohm ``$hc/e$'' conductance oscillations, since they are exponentially sensitive to the polarization of impurity spins. Magnetic field flux $\Phi$ threading  the ring of radius $R$ changes electron wave functions and, consequently, the conductance $g_\Phi$ of the ring.  The conductance
statistics is characterized by the correlation function:
\begin{equation}
\langle\langle
g_{\Phi}g_{\Phi+\Delta \Phi}\rangle\rangle=
\sum_{k=0}^{\infty}  {\cal K}_k \cos \left(2\pi k \frac{\Delta \Phi}{\Phi_0}\right),
\label{eq:1.10}
\end{equation}
where $\Phi=\pi R^2 B$ is the magnetic flux through the ring. 

We find the amplitude
of oscillations of the conductance correlation function:
\begin{eqnarray}
{\cal K}_k\!\! &\!\!  =\!\!  &  \alpha \frac{e^4}{32\hbar^2}\frac{D^{3/2}}{R^3T^2}\int
\frac{e^{-2\pi k R\sqrt{ \Gamma(\e)/D}}}{\sqrt{\Gamma(\e)}}
\frac{d\e}{\cosh^4\!\e\! /2T},
\label{eq:1.11}
\end{eqnarray}
where
\begin{equation}
\Gamma(\varepsilon)=\left[
1-\frac{\langle \hat S_z \rangle^2 + \langle \hat
S_z\rangle\tanh(\e+g\mu_BB)/2T} {S(S+1)}\right]\frac{1}{\tau_{\rm{s}}}
\label{eq:1.12}
\end{equation}
is the phase relaxation rate to the second order in the exchange interaction $J$, and $\alpha$ is a dimensionless geometry dependent factor. As the
magnetic field increases, $\Gamma(\varepsilon)$ decreases exponentially
and at field $B_*$, given by Eq.~(\ref{eq:1.7}), the higher order term in the
exchange interaction, Eq.~(\ref{eq:1.6}), prevails.

Equation~(\ref{eq:1.11}) is valid at $\Gamma(\e\simeq T)\ll T$. This
condition is always satisfied at a sufficiently strong magnetic field, since
$\Gamma(\e\simeq T)\to 0$ as $B$ increases and the impurity spins become
almost polarized, see Eqs.~(\ref{eq:1.6}) and (\ref{eq:1.12}).
Here again we disregarded other mechanisms of phase relaxation, such as the electron-electron interaction.\cite{WL,AB} 

We emphasize that Eq.~(\ref{eq:1.12}) contains the average $\Sodin^2$ instead of the term $\S2$ in Eq.~(\ref{eq:1.5}). The two equations are different because of the nature of the electron dynamics producing the WL correction to the conductivity and conductance fluctuations. The difference can be understood in the following way.

The weak localization correction to the conductivity originates from the electron passage along the same trajectory twice. The time difference between the two passages does not exceed the time of phase relaxation, $[\Gamma (\varepsilon\sim T)]^{-1}$ in our case. In the derivation of Eq.~(\ref{eq:1.5}) we assumed this time being shorter than the Korringa relaxation time\cite{Korringa} $\tau_T$, see Eq.~(\ref{eq:Tkorringa}). Hence the instantaneous spin-spin correlator $\langle S_z^2\rangle$ enters Eq.~(\ref{eq:1.5}). The origin of the correlations in mesoscopic conductance is also due to passage of electrons along the same trajectory. Here, however, the relevant time difference is defined by the time between the measurements, and typically greatly exceeds $\tau_T$. 
That is why the conductance correlation function is described by 
$\langle S_z\rangle^2$ correlator, characterizing a non-fluctuating component of impurity spins.

Note that in the case of low concentration of magnetic impurities 
$n_{\rm s}\lesssim \nu T$ the Korringa relaxation time may become shorter than $[\Gamma (\varepsilon\sim T)]^{-1}$. In this case Eq.~(\ref{eq:1.12}) rather than Eq.~(\ref{eq:1.5}) defines the phase relaxation rate in the WL correction to the conductivity. The corresponding modification of the WL correction at $B=0$ was considered earlier in Ref.~[\onlinecite{meyer}]. 

To summarize, we studied the effect of magnetic field on the weak localization correction and on the magnitude of mesoscopic conductance fluctuations in a conductor with magnetic impurities. Our results are valid at an arbitrary level of the spin polarization. We demonstrate that the electron phase relaxation rate acquires energy dependence due to the Zeeman splitting. Such energy dependence is absent in the limits of no impurity polarization, $B=0$, and strong polarization, $g\mu_BB\gg T$. However, in the experimentally relevant intermediate regime, $g\mu_BB\sim T$, the effect of spin polarization can not be accounted for by assigning a single phase relaxation rate to all electron states.

%%%%%%%%%%%%%%%%%%%%%%%%%%%%%%%%%%%%%%%%%%%%%%%%%%%%%%%%%%%%%%%%%%%%%%%%%%%%%%
%%%%%%%%%%%%%%%%%%%%%%%%%%%%%%%%%%%%%%%%%%%%%%%%%%%%%%%%%%%%%%%%%%%%%%%%%%%%%%
%%%%%%%%%%%%%%%%%%%%%%%%%%%%%%%%%%%%%%%%%%%%%%%%%%%%%%%%%%%%%%%%%%%%%%%%%%%%%%
\section{Model}
\label{sec:3}
%%%%%%%%%%%%%%%%%%%%%%%%%%%%%%%%%%%%%%%%%%%%%%%%%%%%%%%%%%%%%%%%%%%%%%%%%%%%%%
%%%%%%%%%%%%%%%%%%%%%%%%%%%%%%%%%%%%%%%%%%%%%%%%%%%%%%%%%%%%%%%%%%%%%%%%%%%%%%
%%%%%%%%%%%%%%%%%%%%%%%%%%%%%%%%%%%%%%%%%%%%%%%%%%%%%%%%%%%%%%%%%%%%%%%%%%%%%%

We consider a metal with isotropic elastic scattering of electrons by impurities, characterized by the mean scattering rate of electrons $1/\fp = 2\pi\nu n_{\rm e} U^2$.
Here $\nu$ is the electron density of states at the Fermi surface per spin state, 
$n_{\rm e}$ is the concentration of impurities, and $U$ is the Born amplitude of elastic scattering by an impurity.

The scattering of conducting electrons off magnetic impurities is
described by the Hamiltonian: 
\begin{equation}
\hat H_{\rm m}=J \hat{\vec S} \hat {\vec \sigma},
\label{eq:HS} 
\end{equation}
where $\hat{\vec S}$ is the spin operator of
a magnetic impurity, and $J$ is the exchange constant. The electron
scattering rate by magnetic impurities is 
\begin{equation}
\frac{1}{\ts}=2\pi\nu n_{\rm s} J^2 S(S+1).
\label{eq:4b}
\end{equation}
Here $n_{\rm s}$ is the magnetic impurity concentration, and $S$ is the total
impurity spin. The exchange constant $J$ is renormalized
due to the Kondo effect: $J\to 2/(\nu\ln T/T_K)$ at temperatures  $T$ exceeding both $T_K$ and $g\muB B$.
In stronger fields, $B>{\rm max}\{T,T_K\}/g\mu_B$, temperature $T$ under the logarithm is replaced by $g\muB B$.

We study the effect of magnetic impurities on conductivity of metals.
In order to evaluate the interference effects in electron transport we will need instantaneous, $\langle S_z\rangle$ and $\langle S_z^2 \rangle$,
and time delayed spin-spin correlation functions:
\begin{subequations}
\label{eq:5}
\begin{eqnarray}
\chi_{z}(\tau) & = & 
\frac{\langle \hat S_z(t) \hat  S_z(t-\tau)\rangle}{S(S+1)},
\label{eq:5a}
\\
\chi_{\bot}(\tau) & = & 
\frac{\langle [\hat S_+(t), \hat S_-(t-\tau)]_+ \rangle}{S(S+1)}.
\label{eq:5b}
\end{eqnarray}
\end{subequations}
Here $\hat S_z(t)$ and $\hat S_\pm(t)=\hat S_x(t)\pm i \hat S_y(t)$ are
parallel and perpendicular spin components of a magnetic impurity,
$[\cdot,\cdot]_+$ is an anticommutator,  
 and
%\begin{subequations}
\begin{eqnarray}
\langle\hat A\rangle =
\frac{1}{Z}
\sum_{m=-S}^{S}A_{mm}e^{m \os /T},
\label{eq:5_5a}
\ \
Z = \sum_{m=-S}^{S}e^{m \os /T}
\label{eq:5_5b}
\end{eqnarray} 
%\label{eq:5_5}
%\end{subequations}
stands for the thermodynamic average in the presence of a magnetic field  $B$, producing
the Zeeman splitting $\os=g\muB B$ of the magnetic impurity
states; $g$ is the impurity spin gyromagnetic factor.

For a metal with dilute non-interacting magnetic impurities,  functions
$\chi_{z}(\tau)$ and $\chi_{\bot}(\tau)$ have the form:
\begin{subequations}
\label{eq:500}
\begin{eqnarray}
\chi_{z}(\tau) & = & 
\frac{\Sodin ^2+ \left\{\S2 - \Sodin ^2 \right\}f_z(\tau)}
{S(S+1)},
\label{eq:500a}
\\
\chi_{\bot}(\tau) & = & 
\frac{\left\{S(S+1) - \S2 \right\}f_\bot(\tau)} {S(S+1)}.
\label{eq:500b}
\end{eqnarray}
\end{subequations}

The functions $f_z(\tau)$ and $f_\bot(\tau)$ describe spin relaxation. 
In a dilute system of magnetic impurities, relaxation occurs due to the Korringa mechanism\cite{Korringa} and is exponential:
\begin{eqnarray}
f_\bot(\tau)=e^{-|\tau|/\tK_{\bot}+i\os\tau},
\ \ \ \ \ \  
f_z(\tau)=e^{-|\tau|/\tK_{z}}.
\label{relax}
\end{eqnarray}
Here $\tK_{\bot}$ and $\tK_{z}$ are the spin relaxation times for components
perpendicular and parallel to the applied magnetic field, respectively. 
In a weak magnetic field the two
spin relaxation times coincide and are equal to the Korringa time 
$\tau_{\rm T}$:
\begin{equation}
\frac{1}{\tK_{\rm T}}=\frac{1}{\tK_{\bot}}=\frac{1}{\tK_z}
=\frac{2\pi}{3}  (J\nu)^2 T.
\label{eq:Tkorringa}
\end{equation}

The applied magnetic field produces Zeeman splitting 
$\varepsilon_{\rm Z}$ of energy states of transport electrons. 
When estimating the Zeeman energy $\varepsilon_Z$ in metals with magnetic impurities, it is important to notice
that the sign of the exchange constant $J$ is fixed by the nature
of the pair of host and impurity~\cite{jarozunski} atoms. Polarization of
magnetic impurities results in the exchange contribution
$n_{\rm s} J\langle S_z\rangle/\muB$ to the effective magnetic
field which causes the Zeeman splitting $\varepsilon_{\rm Z}$ of  spin
states of transport electrons with the gyromagnetic factor $g_{\rm e}$
\begin{equation}
\varepsilon_{\rm Z}=g_{\rm e} \muB B- 2n_{\rm s} J \Sodin .
\label{zeeman}
\end{equation}
Equation (\ref{zeeman}) shows that magnetic impurities may
significantly affect  the Zeeman splitting of transport electron states. 

In the case of antiferromagnetic exchange, $J>0$, the dependence of
$\varepsilon_Z$ on $B$ is not monotonic at low temperatures. As we
will see below, Eq.~(\ref{zeeman}) provides a mechanism for a
non-monotonic in $B$ interference contribution to the metal
conductivity, similar to the Jaccarino-Peter mechanism of the
reentrant superconductivity.\cite{Jaccarino,Jaccarino-exp}

We also consider the effect of spin-orbit scattering on the conduction of electric current.
For the specific case of heavy element impurities of concentration
$n_{\rm so}$, the corresponding term of the Hamiltonian has the
form:\cite{SO} $\hat H_{\rm so}=U_{\rm so}[\p\times\ps]\hat {\vec
  \sigma}$, and the spin-orbit scattering rate $\gamma_{\rm so}
=2\pi\nu n_{\rm so} U^2_{\rm so}p_{\rm F}^4$.

In this paper we assume that the shortest electron relaxation time is due to elastic scattering and calculate the conductivity to the lowest order in
$\fp\gs$ and $\fp\gso$, using the standard diagrammatic technique for
a disordered metal.\cite{AGD} Nevertheless the results are valid even
for conductors with strong spin-orbit interaction in their host
material, when the spin-orbit scattering rate is comparable with
elastic scattering rate.

%%%%%%%%%%%%%%%%%%%%%%%%%%%%%%%%%%%%%%%%%%%%%%%%%%%%%%%%%%%%%%%%%%%%%%%%%%%%%%%
%%%%%%%%%%%%%%%%%%%%%%%%%%%%%%%%%%%%%%%%%%%%%%%%%%%%%%%%%%%%%%%%%%%%%%%%%%%%%%%
%%%%%%%%%%%%%%%%%%%%%%%%%%%%%%%%%%%%%%%%%%%%%%%%%%%%%%%%%%%%%%%%%%%%%%%%%%%%%%%
\section{Weak localization correction to the conductivity}
\label{sec:4}
%%%%%%%%%%%%%%%%%%%%%%%%%%%%%%%%%%%%%%%%%%%%%%%%%%%%%%%%%%%%%%%%%%%%%%%%%%%%%%%
%%%%%%%%%%%%%%%%%%%%%%%%%%%%%%%%%%%%%%%%%%%%%%%%%%%%%%%%%%%%%%%%%%%%%%%%%%%%%%%
%%%%%%%%%%%%%%%%%%%%%%%%%%%%%%%%%%%%%%%%%%%%%%%%%%%%%%%%%%%%%%%%%%%%%%%%%%%%%%%

The weak localization correction to the conductivity is given by the
following expression:
\begin{eqnarray}
\Delta \sigma({\bf B}) & = & 
-\frac{e^2D}{\pi\hbar}
\int \frac{d\e}{4T\cosh^2\e/2T}
\nonumber
\\
&& \times \int\frac{d^d\q}{(2\pi)^d}
\sum_{i=1}^{4}
\alpha_i{\cal C}_i(\e,\o=0,\q),
\label{eq:11a}
\end{eqnarray}
where the Cooperon ${\cal C}(\e,\o,\q)$ is 
\begin{equation}
{\cal C}_i(\e,\o,\q)=
\frac{1}{i\o+D\q^2+\Gamma_i(\e)}.
\label{eq:11}
\end{equation}
The Cooperon components, corresponding to different indices $i$, represent
possible spin configurations: indices $i=1,2$ are assigned to the Cooperon
components with non-zero spin projections on the direction
of magnetic field, $m_{\rm c}=\pm 1$ and $S_{\rm c}=1$; index $i=3$ is assigned to the Cooperon with $S_{\rm c}=1$ and zero spin projection 
$m_{\rm c}=0$; and index $i=4$ stands for a singlet component, 
$S_{\rm c}=m_{\rm c}=0$, see Table~\ref{tab:1}.

Performing integration over momentum $\q$, we obtain a general result in
the form
\begin{equation}
\label{eq:12}
\Delta \sigma = -\frac{e^2}{\pi \hbar}\sum_{i=1}^{4}\alpha_i
{\cal F}_{d}({\bf B},\Gamma_i).
\end{equation}
Functions ${\cal F}_{d}({\bf B},\Gamma)$ have different form depending on the  conductor geometry ($d=2$ for a metal film and $d=1$ for a wire), while
parameters $\alpha_{i}$ and phase relaxation rates $\Gamma_{i}(B)$ are
determined by details of electron scattering processes. First we present
functions ${\cal F}_{d}({\bf B},\Gamma)$ 
for different conductor geometries and
then discuss the effect of spin flip scattering on the weak
localization correction to the conductivity.

%%%%%%%%%%%%%%%%%%%%%%%%%%%%%%%%%%%%%%%%%%%%%%%%%%%%%%%%%%%%%%%%%%%%%%%%%%%%%%%
%%%%%%%%%%%%%%%%%%%%%%%%%%%%%%%%%%%%%%%%%%%%%%%%%%%%%%%%%%%%%%%%%%%%%%%%%%%%%%%
\subsection{Geometry dependence}
\label{sec:4.1}
%%%%%%%%%%%%%%%%%%%%%%%%%%%%%%%%%%%%%%%%%%%%%%%%%%%%%%%%%%%%%%%%%%%%%%%%%%%%%%%
%%%%%%%%%%%%%%%%%%%%%%%%%%%%%%%%%%%%%%%%%%%%%%%%%%%%%%%%%%%%%%%%%%%%%%%%%%%%%%%
The weak localization correction to the conductivity of a quasi two
dimensional metal is determined by the following expression:
\begin{eqnarray}
{\cal F}_{2}\left({\bf B},\Gamma_i\right)  = & - &\frac{1}{4\pi}\left[
\psi\left(\frac{1}{2}+\frac{\Gamma_i +\gH}{4DeB_\bot/\hbar c}\right)\right.
\nonumber
\\
&  &\mbox{} +  \left. %\frac{1}{4\pi}
%+
\ln\frac{4eB_\bot D \fp}{\hbar c }\right]
,
\label{eq:13}
\end{eqnarray}
where $\psi(x)$ is the digamma function and $\Gamma_i$ is the
dephasing rate for the $i$th component of the Cooperon. Here
$\Gamma_i$ includes dephasing $\gH$ by the applied magnetic
field parallel to the film $\gH=e^2B_\|^2 Da^2/12 \hbar^2c^2$, where $B_\bot$
and $B_\|$ are the magnetic field components, perpendicular and
parallel to the film. In weak perpendicular magnetic
field we use $\psi(1/2+x)\approx\ln x+1/24x^2$ and obtain:
\begin{eqnarray}
{\cal F}_{2}\left({\bf B},\Gamma_i\right) & = &\frac{1}{4\pi}
\left[\ln \frac{1}{(\Gamma_i+\gH)\fp } 
\right.
\nonumber
\\
& - &\left. \frac{2(DeB_\bot/\hbar c)^2}{3(\Gamma_i+\gH)^2}
\right].
\label{eq:14}
\end{eqnarray}
The terms with $\fp$ in Eqs.~(\ref{eq:13}) and (\ref{eq:14})
originate from the ultraviolet logarithmic divergence of the momentum integral
for the weak localization correction to the conductivity in two dimensional
case. The existence of this term complicates analysis of the
magnetoresistance measurements. 

In the quasi one dimensional case the integral over momentum gives:
\begin{equation}
\label{eq:15}
{\cal F}_{1}\left({\bf B},\Gamma_i\right)=\frac{1}{2}
\sqrt{\frac{D}{\Gamma_i+\gH}}.
\end{equation}
The denominator of Eq.~(\ref{eq:15}) contains the sum of electron phase relaxation rates due to scattering on magnetic impurities and due to the orbital effect of the applied magnetic field $\gH$. For a magnetic field of strength $B$ and applied in direction $\n$, the orbital contribution to the phase relaxation rate is
\begin{eqnarray}
\label{eq:16}
\gH= \vartheta (\n) \frac{e^2B^2 D A}{\hbar^2c^2} ,
\end{eqnarray}
where the function $\vartheta (\n)$ of the direction $\n$
is of the order of unity, and $A$ is the area of
the wire cross-section.

\begin{table}
\caption{The Cooperon phase relaxation factors for various components corresponding different spin configurations. 
The components with $i=1,2,3$ represent a spin states with total spin one,
$S_{\rm C}=1$. These spin configurations form a triplet in an isotropic system. Anisotropy due to    
the applied magnetic field splits the degeneracy. The $i=4$ component is a 
singlet spin configuration with zero total momentum, $S_{\rm C}=0$.  } 
\begin{ruledtabular}
\begin{tabular}{|c|ccc|}
$\ i\ $ & $|S_{\rm C},m_{\rm C}\rangle $ & $\Gamma_{i}\ts $ & $\alpha_{i}\ $ \\
\hline 
$\ 1\ $ & $S_{\rm C}=1, m_{\rm C}=+ 1$ 
& $ 1-\S2/S^2$ & $ +1\ $ \\ 
$  2  $ & $S_{\rm C}=1, m_{\rm C}=- 1$ 
& $1-\S2/S^2$ & $ +1\ $ \\ 
$  3  $ & $S_{\rm C}=1, m_{\rm C}=\mbox{ }0$   
& $2 \S2/S^2   $ & $ +1\ $ \\ 
$  4  $ & $S_{\rm C}=0, m_{\rm C}=\mbox{ }0$   
& $2        $ & $-1\ $ \\
\end{tabular}
\end{ruledtabular}
\label{tab:1}
\end{table}

In the following subsections we focus on properties of the
phase relaxation rate $\Gamma_i$ due to scattering on magnetic impurities.  To be specific, we will present the WL correction to the conductivity 
of a  thin wire. Since the phase relaxation factors $\Gamma_i$ do not depend on the sample geometry, the results for the WL correction to the conductivity of the next two subsections can be easily generalized to other system geometries, such as a film, a metal ring or an open quantum dot.\cite{qd}
 
%%%%%%%%%%%%%%%%%%%%%%%%%%%%%%%%%%%%%%%%%%%%%%%%%%%%%%%%%%%%%%%%%%%%%%%%%%%%%%
%%%%%%%%%%%%%%%%%%%%%%%%%%%%%%%%%%%%%%%%%%%%%%%%%%%%%%%%%%%%%%%%%%%%%%%%%%%%%%
%%%%%%%%%%%%%%%%%%%%%%%%%%%%%%%%%%%%%%%%%%%%%%%%%%%%%%%%%%%%%%%%%%%%%%%%%%%%%%
\subsection{Effect of classical spin impurities}
\label{sec:4.2}
%%%%%%%%%%%%%%%%%%%%%%%%%%%%%%%%%%%%%%%%%%%%%%%%%%%%%%%%%%%%%%%%%%%%%%%%%%%%%%
%%%%%%%%%%%%%%%%%%%%%%%%%%%%%%%%%%%%%%%%%%%%%%%%%%%%%%%%%%%%%%%%%%%%%%%%%%%%%%
%%%%%%%%%%%%%%%%%%%%%%%%%%%%%%%%%%%%%%%%%%%%%%%%%%%%%%%%%%%%%%%%%%%%%%%%%%%%%%
Even in the absence of spin-orbit interaction, there are five parameters affecting the WL correction to the conductivity, including system temperature $T$, the applied magnetic field $B$, impurity spin relaxation time $\tau_{\rm T}$, the electron scattering rate on magnetic impurities $1/\ts$, and the phase relaxation rate due to the orbital effect of the applied magnetic field $\gH$. It is easier to establish the role of these parameters in the case of classical ($S\gg 1$) spins, considered in this Subsection.

\begin{figure}
\centerline{\epsfxsize=8cm
\epsfbox{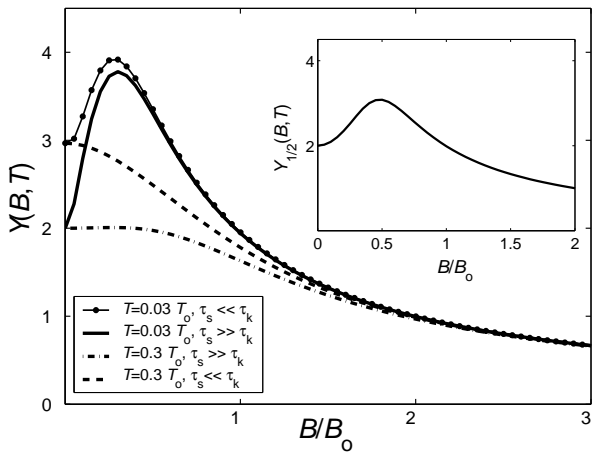}}
\caption {
Dependence of the weak localization correction on the applied magnetic
field in a metal with classical spin impurities. The plot shows function
$Y(B,T)$ at two temperature values $T=0.3T_{\rm o}$ and $T=0.03T_{\rm o}$
for both short ($\tau_{\rm T}\ll\tau_{\rm s}$) and long ($\tau_{\rm
K}\gg\tau_{\rm s}$) impurity spin relaxation time $\tau_{\rm T}$. The subplot shows the WL correction to the conductivity for $S=1/2$ impurities at $T=T_{\rm o}$. Here $T_{\rm o}$ is given by Eq.~(\ref{eq:20}). }
\label{fig:wlplot}
\end{figure}

The explicit matrix form of the equation for the Cooperon in the case of classical spins is given in Appendix~\ref{app:A}, see Eqs.~(\ref{eq:9}) therein. For an arbitrary relation between the electron dephasing rate $\Gamma_{i}$ and the impurity spin relaxation rate $1/\tau_{\rm T}$, the solution of this equation is cumbersome.
Here we consider only the limiting cases of long 
($\tau_{\rm T}\Gamma_{i}\gg 1$) and
short ($\tau_{\rm T}\Gamma_{i}\ll 1$) spin relaxation time. Usually (for not too small concentration of magnetic impurities 
$n_{\rm s}\gtrsim \nu T$) the first condition is satisfied; the second condition may become relevant only at very low $n_{\rm s}$.
In the first case one may neglect\cite{HLN} the dynamics of localized spins setting $\tau=0$ in the correlation functions  Eq.~(\ref{eq:9b}). In the second case, we need to set $\tau\to\infty$ in Eq.~(\ref{eq:9b}). 
In both limits we obtain the following expressions for the parameters of Eq.~(\ref{eq:12})
\begin{equation}
\begin{array}{ccccc}
\alpha_{1,2}=1, &&
\displaystyle
\alpha_{3}=+
\frac{\chi_\bot}{Z\ts},
&&
\displaystyle
\alpha_{4}=-
\frac{\chi_\bot}{Z\ts}
\end{array}
\label{eq:17alpha}
\end{equation}
and for the phase relaxation rates
\begin{subequations}
\label{eq:18}
\begin{eqnarray}
\Gamma_{1,2} & = & (1-\chi_{z})\frac{1}{\ts},
\label{eq:18a}
\\
\Gamma_{3} & = & (1+\chi_{z})\frac{1}{\ts}+i\os - Z,
\label{eq:18b}
\\
\Gamma_{4} & = & (1+\chi_{z})\frac{1}{\ts}+i\os + Z.
\label{eq:18c}
\end{eqnarray}
\label{eq:17}
\end{subequations}
Here
$$
Z=\sqrt{\frac{\chi^2_\bot }{\ts^2}-(\e_{\rm Z}+\os)^2};
$$
$\chi_{z}$ and $\chi_\bot$ are the impurity spin correlators given by
Eqs.~(\ref{eq:500}) and (\ref{relax}) with proper substitution $\tau=0$ or $\tau\to \infty$. The correlation functions $\Sodin$ and $\S2$ of classical spins are described by the following functions:
\begin{subequations}
\begin{eqnarray}
\langle S_z \rangle & = &
S\coth\frac{S\os}{T}-\frac{T}{\os},
\label{eq:6a}
\\
\langle S^2_z \rangle
 & = &  S^2-\frac{2T}{S\os}\coth\frac{S\os}{T}+
2\left(\frac{T}{\os}\right)^2\!\!\! .
\label{eq:6b}
\end{eqnarray}
\label{eq:6_cl}
\end{subequations}

The weak localization correction to the conductivity of a wire may be represented as
\begin{eqnarray}
\Delta\sigma_1  = 
-\frac{e^2}{2\pi\hbar}\sqrt{D\tau_{\rm s}}\ 
Y \left( {\bf B},T\right).
\label{eq:22}
\end{eqnarray}
The function $Y({\bf B}, T)$ has different forms, depending on the
relations between parameters of the system. 

In the limit of long spin relaxation time ($\tau_{\rm T}\Gamma_i\gg 1$) and 
weak magnetic field ($\os\ts \ll 1$ and $\e_{\rm Z} \ts \ll 1$) all four Cooperon modes contribute to the weak localization correction to the conductivity
\begin{eqnarray} 
\!\!\!
Y\left( {\bf B}, T\right)\!\! \!\! &=&\!\! \!\!
\frac{2}{\sqrt{1-\S2/S^2 + \gH\tau_{\rm s}}}
\nonumber
\\
& \!\!+\!\! &\!\!
\frac{1}{\sqrt{2 \S2/S^2 +\gH\tau_{\rm s} }}\! -
\! \frac{1}{\sqrt{2 + \gH\tau_{\rm s} }}.
\label{eq:22a}
\end{eqnarray}
The first term in Eq.~(\ref{eq:22a}) is due to the Cooperon modes with
$i=1,2$. It is the first term of the function $Y(B,T)$ which is responsible for a non-monotonic magnetic field dependence of the WL correction to the conductivity. As magnetic field increases and impurity spins become polarized, the phase relaxation rate for modes $i=1,2$ decreases, and the first term in Eq.~(\ref{eq:22a}) grows. 
The combination of the remaining two terms, representing  Cooperon components $i=3,4$, constitutes  $0.17 \Delta\sigma_1$
of the  full value of $\Delta\sigma_1$ at zero magnetic field and monotonically decreases as the magnetic field increases.

We notice that the contribution of $i=3,4$ Cooperon components diminishes not only as magnetic impurities become polarized, but also as the Zeeman splitting of conduction electrons and impurity spins ($\e_{\rm Z}+\os$) increases, see Eqs.~(\ref{eq:11a}) and (\ref{eq:17alpha}). 
At strong magnetic field, $|\os+\e_{\rm Z}|\ts \gg 1$, 
the contribution  of $i=3,4$ Cooperon components to $\Delta\sigma_1$  vanishes, since both the Zeeman splitting of the conduction electrons and the energy transferred in the spin flip process destroy the interference of time reversed paths in the Cooperon.
If the spin relaxation time is still long, $\tau_{\rm T}\Gamma_i\gg 1$,
we have
\begin{equation} 
Y\left( {\bf B}, T\right) =
\frac{2}{\sqrt{1-\S2/S^2 + \gH\tau_{\rm s}}}.
\label{eq:22highT}
\end{equation}

For impurity spin system with short spin relaxation time, 
$\tau_{\rm T}\Gamma_i\ll 1$, the Cooperon modes with $i=3,4$ do not contribute to $\Delta\sigma_1$ again. In this case, the weak localization correction to the conductivity  is given by Eq.~(\ref{eq:22}) with
\begin{equation} 
Y\left( {\bf B}, T\right) =
\frac{2}{\sqrt{1-\Sodin ^2/S^2 + \gH\tau_{\rm s}}}
.
\label{eq:22b}
\end{equation}
We notice that as magnetic field increases, the phase relaxation rate decreases, and the condition of short spin relaxation time 
($\tau_{\rm T}\Gamma_{1,2}\ll 1$) may be reached, even though initially system was in the opposite regime, $\tau_{\rm T}\gg \ts$.

We conclude from Eqs.~(\ref{eq:22a})--(\ref{eq:22b}) that the WL correction to the conductivity depends on the polarization of magnetic impurities through the thermodynamic averages $\Sodin $ and $\S2$. The polarization is significant in an applied field of the order of $B = T/g\muB S$, shown as a straight solid line in Fig.~\ref{fig:TB-plane}.

The orbital effect of the applied magnetic field is represented by the term $\gH\tau_{\rm s}$ in Eqs.~(\ref{eq:22a})--(\ref{eq:22b}). The orbital  effect and the impurity spin polarization compete with each other: the orbital effect suppresses, while the spin polarization enhances the WL correction to the conductivity.  The phase relaxation rates  $\gH$ and $\Gamma_1$ become equal, $\gH=\Gamma_1$, at magnetic field  $B_{\rm o}(T)$:
\begin{equation}
\label{eq:19}
B_{\rm o}(T) = \frac{\Phi_0}{\sqrt{D\ts A}}
\sqrt{1-\frac{\Sodin ^2_{B=B_{\rm o}}}{S^2}}.
\end{equation}
The effect of spin scattering prevails over orbital effect of the magnetic field at $B\lesssim B_{\rm o}(T)$. On the other hand, the enhancement of the WL correction due to impurity spin polarization is significant at 
$B\gtrsim T/g\muB S$.  These two limitations are met if the wire temperature $T\lesssim T_{\rm o}$, where
\begin{equation}
\label{eq:20}
T_{\rm o}=  g \muB S\frac{\Phi_0}{\sqrt{D\ts A}}.
\end{equation}
At higher temperature,  $T\gtrsim T_{\rm o}$, the effect of the spin polarization is concealed by the orbital effect.

In Fig.~\ref{fig:wlplot} we illustrate the effect of an applied magnetic field on the WL correction to the conductivity. We show dependence of the function $Y(B,T)$ on the applied magnetic field $B$ at temperatures $T=0.3T_{\rm o}$ and $T=0.03T_{\rm o}$. At low temperature $T=0.03T_{\rm o}$, the effect of
impurity spin polarization is significant, but already at temperature
$T=0.3T_{\rm o}$ it fades away.

The spin-orbit interaction modifies the WL correction to the conductivity.
Using Eqs.~(\ref{eq:9}) in Appendix~\ref{app:A} we find:
\begin{subequations}
\label{eq:17alphaSO}
\begin{eqnarray}
\alpha_{1,2} & = & 1,
\\
\displaystyle
\alpha_{3} & = &
\frac{2\gso/3-\chi_\bot/\ts}{\sqrt{(2\gso/3-\chi_\bot/\ts)^2-\e_{\rm Z}^2}},
\\
\displaystyle
\alpha_{4} & = & -
\frac{2\gso/3-\chi_\bot/\ts}{\sqrt{(2\gso/3-\chi_\bot/\ts)^2-\e_{\rm Z}^2}}
\end{eqnarray}
\end{subequations}
and the phase relaxation rates are
\begin{subequations}
\label{eq:18SO}
\begin{eqnarray}
\!\!\!\!\Gamma_{1,2}\!\! & = &\!\! (1-\chi_{zz})\frac{1}{\ts}
+\!\!\frac{4\gso}{3},
\label{eq:18aSO}
\\
\!\!\!\!\Gamma_{3}\!\! & = & \!\! \frac{1+\chi_{zz}}{\ts}\!\!+\!\!\frac{2\gso}{3}+\!\!
\sqrt{\left|\frac{\chi_\bot}{\ts}-\frac{2\gso}{3}\right|^2-\e_{\rm Z}^2},
\label{eq:18bSO}
\\
\!\!\!\!\Gamma_{4}\!\! & = & \!\! \frac{1+\chi_{zz}}{\ts}\!\!+\!\!\frac{2\gso}{3}-\!\!
\sqrt{\left|\frac{\chi_\bot}{\ts}-\frac{2\gso}{3}\right|^2-\e_{\rm Z}^2}.
\label{eq:18cSO}
\end{eqnarray}
\label{eq:17SO}
\end{subequations}
Here we assumed $\os\ll \Gamma_{i}$ for simplicity.

According to Eqs.~(\ref{eq:18SO}), in metals with strong spin-orbit scattering, $\gso \ts \gg 1$, the terms with $i=1,2,3$ are suppressed and the interference correction to the conductivity is
described by the singlet antilocalization term $i=4$. In the limit of long
impurity spin relaxation time $\tau_{\rm T}\gg \ts$, the weak localization correction to the conductivity depends on $\tau_{\rm s}$ and the Zeeman energy $\varepsilon_Z$:
\begin{eqnarray}
Y({\bf B},T) & = &
-\frac{1}{\sqrt{2+3\e_{\rm Z}^2\tau_{\rm s}/4\gso+
\gH\tau_{\rm s}}} ,
\label{wlso}
\end{eqnarray}
while in the opposite limit $\tau_{\rm T}\ll \ts$ the correction also depends on the polarization of impurity spins:
\begin{eqnarray}
Y({\bf B},T) = 
-\frac{1}
{\sqrt{1+ \Sodin ^2/S^2 + 
3\e_{\rm Z}^2\tau_{\rm s}/4\gso+\gH\tau_{\rm s}}}.
\label{wlso1}
\end{eqnarray}

The terms $\gH\ts \propto A$ in the denominator of Eqs.~(\ref{wlso})
and (\ref{wlso1}) originate from the orbital magnetic field effect, see Eq.~(\ref{eq:16}). If the area $A$ is
sufficiently small so that the orbital part is not important, and the
temperature is low enough, then the antilocalization may show a
non-monotonic field dependence due to the  Jaccarino-Peter
mechanism,\cite{Jaccarino}  see Eq.~(\ref{zeeman}).

\begin{figure*}
\centerline{\epsfxsize=11cm
\epsfbox{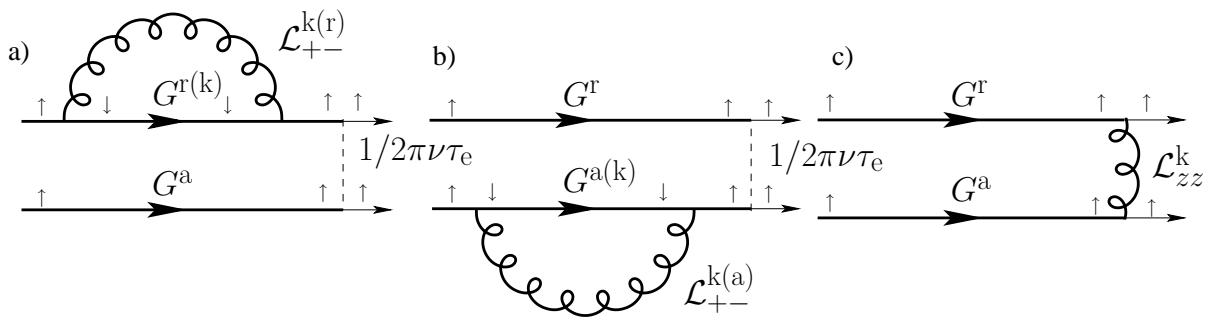}}
\caption {
The Cooperon decay rate is determined by diagrams of the second
order in the exchange constant $J$. The two left
diagrams represent the electron self energy contribution from processes with and without (not shown) spin flip. The right diagram represents the vertex
correction.}
\label{fig:Cse}
\end{figure*}

From the above analysis performed in the approximation of large
impurity spins we make two conclusions. First, the polarization of
impurity spins hardly affects the weak (anti)localization correction
to the conductivity in metals with strong spin-orbit interaction.
Indeed, the only surviving contribution to the conductivity correction
originates from the singlet spin configuration of the Cooperon, which
is not sensitive to the impurity spin polarization. Second, in metals
without spin-orbit interaction, the Cooperon modes with zero spin
projection on the magnetic field ($i=3,4$) give only a small ($\sim
0.17\Delta\sigma$) contribution to the total WL correction to the
conductivity, and their contribution vanishes at moderately strong
($\os\ts\sim 1$) magnetic field.  In the next subsection we neglect
these terms and calculate the WL correction to the conductivity,
originating from Cooperon modes $i=1,2$ at such fields that
$\os\ts\gtrsim 1$. We perform the calculations for an arbitrary value
of $S$.

%%%%%%%%%%%%%%%%%%%%%%%%%%%%%%%%%%%%%%%%%%%%%%%%%%%%%%%%%%%%%%%%%%%%%%%%%%%%%%
%%%%%%%%%%%%%%%%%%%%%%%%%%%%%%%%%%%%%%%%%%%%%%%%%%%%%%%%%%%%%%%%%%%%%%%%%%%%%%
%%%%%%%%%%%%%%%%%%%%%%%%%%%%%%%%%%%%%%%%%%%%%%%%%%%%%%%%%%%%%%%%%%%%%%%%%%%%%%
\subsection{Effect of quantum spin impurities}
\label{sec:4.3}
%%%%%%%%%%%%%%%%%%%%%%%%%%%%%%%%%%%%%%%%%%%%%%%%%%%%%%%%%%%%%%%%%%%%%%%%%%%%%%
%%%%%%%%%%%%%%%%%%%%%%%%%%%%%%%%%%%%%%%%%%%%%%%%%%%%%%%%%%%%%%%%%%%%%%%%%%%%%%
%%%%%%%%%%%%%%%%%%%%%%%%%%%%%%%%%%%%%%%%%%%%%%%%%%%%%%%%%%%%%%%%%%%%%%%%%%%%%%
We notice that the semiclassical description is applicable only
\textit{i)} for a large spin $S\gg 1$, when linear in $S$ contributions to
the dephasing rates can be neglected and \textit{ii)} at high temperature
$T\gg \os$, so that the discreetness of spin energy levels can be
disregarded. In this Subsection we consider quantum spins with arbitrary value of $S$ and for arbitrary ratio of $\os/T$. 
As we discussed in the previous Subsection,  only
components $i=1,2$ of the Cooperon with parallel spins, see Table \ref{tab:1}, are important at strong magnetic field.
The terms with $i=3,4$ are small at relatively strong magnetic fields, 
$|\os+\e_{\rm Z}|\tau_s\gtrsim 1$. Moreover, even at weak fields,  
modes $i=3,4$ give rise only to a small portion of $\Delta\sigma_1$, equal to $0.17 \Delta\sigma_1$ at zero magnetic field and monotonically decreasing at stronger fields $B$. To avoid cumbersome expressions, we omit the terms with $i=3,4$ in our further analysis.

The contribution to the Cooperon self energy due to the scattering off
magnetic impurities is shown in Fig.~\ref{fig:Cse}.  Details of our calculations are presented
in Appendix \ref{app:B}. We find that the WL correction to the conductivity
may be represented in the form of Eq.~(\ref{eq:22})
with function $Y({\bf B},T)$ replaced by
\begin{equation}
Y_S({\bf B},T) = 
\int\limits_{-\infty}^{+\infty}
\frac{d\e}{2T\cosh^2\e/2T}
\frac{1}{\sqrt{(\Gamma(\e) + \gH)\ts}},
\label{eq:1.4rY}
\end{equation}
where the phase relaxation rate due to scattering off magnetic impurities is
\begin{equation}
\Gamma(\varepsilon)
=
\left[
1 -
\frac{\langle \hat{S}_{z}^{2}\rangle +
\langle \hat{S}_{z}\rangle \tanh(\varepsilon+\omega_{\rm{s}})/2T
}{S(S+1)}
\right] \frac{1}{\ts},
\label{eq:cse_rate}
\end{equation}
provided that the impurity spin relaxation time is large, 
$\tau_{\rm T}\Gamma(\e\sim T)\gg 1$.

Spin correlation functions $\langle \hat S_z \rangle$ and $\langle \hat
S^2_z \rangle$ in Eq.~(\ref{eq:cse_rate}) are defined by Eqs.~(\ref{eq:5_5b}).
We notice that in strong fields $\os \gtrsim T$ the impurity spins are polarized and the phase relaxation rate of 
electrons close to the Fermi surface $|\e|\lesssim T$ vanishes, in agreement with Ref.~[\onlinecite{BFK}].

We consider the special case of magnetic impurities with $S=1/2$ in more details. First, we discuss the meaning of  the energy-dependent relaxation rate $\Gamma(\e)$,  Eq.~(\ref{eq:cse_rate}). 
In the $S=1/2$ case, $\Gamma(\e)$ can be rewritten in the form:
\begin{equation}
\Gamma(\varepsilon)= \frac{4}{3\ts}
\left[
p_\downarrow (1-n(\e+\os))+p_\uparrow n(\e+\os)
\right] ,
\label{eq:cse1:2}
\end{equation}
where $p_{\uparrow(\downarrow)} = (2\cosh \os/2T)^{-1}\exp{(\pm \os/2T)}$ is the
probability for the spin impurity to be parallel (antiparallel) to the
direction of the magnetic field and $n(\e)=[1+\exp(\e/T)]^{-1}$ is the Fermi
occupation number for electrons with energy $\e$ at temperature $T$. 
We interpret Eq.~(\ref{eq:cse1:2}) in the following way. Two processes contribute to the electron phase relaxation rate: 
\textit{i)} an electron
with spin up and energy $\e$ is scattered by an impurity with spin down to
the electron state with spin down and energy $\e+\os$;
\textit{ii)} a hole with spin up and energy
$\e$ is scattered by an impurity with spin up to the electron state with spin down
and energy $\e+\os$.  The probabilities of these
processes are determined by the first and second terms in
Eq.~(\ref{eq:cse1:2}), respectively.

In the limit of short impurity spin relaxation time, 
$\tau_{\rm T}\Gamma(\e\simeq T) \gg 1$, the function $Y_{ ^1\! /\! _2}({\bf B},T)$ has the form 
\begin{widetext}
\begin{eqnarray}
Y_{ ^1\! /\! _2}({\bf B},T) &  = & 
\int\limits_{-\infty}^{+\infty}  \frac{d x}{\cosh^2 x}
\left[
\frac{2\cosh x}
{3\cosh \os/2T\cosh(x+\os/2T)}
+\gH\tau_{\rm s}\right]^{-1/2}.
\label{eq:wl1:2}
\end{eqnarray} 
\end{widetext}

We illustrate the behavior of function $Y_{ ^1\! /\! _2}(B,T)$ in the inset of Fig.~\ref{fig:wlplot} at temperature $T=0.3T_{\rm o}$, 
where $T_{\rm o}$ is defined by Eq.~(\ref{eq:20}) with $S=1/2$.

If the orbital effect of the magnetic field is small, $\gH=0$, we have
\begin{equation}
Y_{ ^1\! /\! _2}({\bf B},T)=\sqrt{\frac{8}{3}}
\frac{\sinh(3\os/4T)}{\tanh(\os/2T)\sqrt{\cosh(\os/2T)}},
\label{eq:Y-12}
\end{equation}
and the weak localization correction to the conductivity is given by Eq.~(\ref{eq:22}) with $Y_{ ^1\! /\! _2}({\bf B},T)$ in the form of Eq.~(\ref{eq:Y-12}). 
In the opposite limit, when  the orbital effect of the magnetic field on WL dominates over the effect of spin scattering, we may expand function
$Y_{ ^1\! /\! _2}(B,T)$ in $(\ts\gH)^{-1}$ and obtain 
\begin{widetext}
\begin{eqnarray}
\Delta\sigma_1(B) & = & - \frac{e^2}{\pi\hbar}
\sqrt{\frac{D}{\gH}}
\left(1-\frac{1}{3}\frac{1}{\tau_{\rm s}\gH}
\frac{\os}{T\sinh \os/T}\right).
\label{eq:wl1:2b}
\end{eqnarray} 
We notice that the second term in Eq.~(\ref{eq:wl1:2b}) has the same structure as the phase relaxation rate suggested in Ref.~[\onlinecite{meyer}] to describe the WL correction to the conductivity. 
Therefore, the expansion, presented in Eq.~(\ref{eq:wl1:2b}), establishes the conditions of applicability of the suggested formula.\cite{meyer}

Expression for the phase relaxation rate in Eq.~(\ref{eq:cse_rate}) was derived in the limit of long impurity spin relaxation time, 
$\tau_{\rm T}\Gamma(\e\sim T)\gg 1$. In the opposite limit, the spins at different moments of time are not correlated, and we have to substitute $\Sodin^2$ for $\S2$ in Eq.~(\ref{eq:cse_rate}). Then, the function  
$Y_{ ^1\! /\! _2}({\bf B},T)$ acquires the form: 
\begin{equation}
Y_{ ^1\! /\! _2}(B,T)=\int \left( 1-\frac{1}{3}\tanh^2\frac{\os}{2T}-
\frac{2}{3}\tanh\left(x+\frac{\os}{2T}\right)\tanh\frac{\os}{2T}
+\gH\ts
\right)^{-1/2}
\frac{d x}{\cosh^2 x}
\label{eq:wl1:2f}.
\end{equation}
\end{widetext}

We observe from Eq.~(\ref{eq:cse_rate}), that at strong magnetic field the 
phase relaxation rate exponentially vanishes. Consequently, higher order terms in the exchange constant $J$ may become important. The fourth order diagrams for the Cooperon phase relaxation rate are shown in Fig.~\ref{fig:RPA} and the corresponding analytical calculations, presented in Appendix \ref{app:C}, yield
\begin{equation}
\Gamma(\e)=\frac{1}{2\pi} 
\frac{\nu}{n_{\rm s}\ts} \frac{\pi^2T^2+\e^2}{\os^2(S+1)^2} \frac{1}{\ts}.
\label{eq:SE2}
\end{equation}  
We emphasize that  Eq.~(\ref{eq:SE2}) represents the $J^4$ contribution to the electron phase relaxation rate, while Eq.~(\ref{eq:cse_rate}) represents the $J^2$ contribution. The corresponding small parameter of the expansion in powers of $J^2$ may be written as $\nu/(n_{\rm s}\ts)$.

As the applied magnetic field increases, both the $J^2$ and $J^4$ contributions may become comparable, since the $J^2$ contribution decreases exponentially, while the $J^4$ contribution at energy $|\e|\sim T$ decreases much slower, cf. Eqs.~(\ref{eq:cse_rate}) and (\ref{eq:SE2}).
The  small factor $\nu/(n_{\rm s}\ts)$ is compensated at strong enough magnetic field $B_*$, which we estimate from $\exp(-\os/T)=\nu/(n_{\rm s}\ts)$.  We obtain
\begin{equation}
B_*(T)= \frac{T}{g\muB}\ln \frac{\nu}{n_{\rm s}\ts}.
\label{eq:Bstar}
\end{equation}

At small fields, $B\lesssim B_*$, the relaxation rate decreases exponentially with the increase of $B$, whereas at higher fields this dependence is replaced by a slower power law $\Gamma(\varepsilon\sim T)\propto T^2/\os^2$. 
The counterpart of Eq.~(\ref{eq:wl1:2b}) at $B>B_*$ reads
\begin{eqnarray}
\Delta\sigma_1(B) & = & -\frac{1}{\pi^2} \frac{e^2}{\hbar c}
\sqrt{\frac{D}{\gH}}
\nonumber
\\
&\times &
\left(1-
\frac{\pi}{3}\frac{\nu}{n_{\rm s} \tau_{\rm s}}\frac{1}{\ts\gH}
\frac{T^2}{(S+1)^2\os^2}
\right).
\label{eq:wl1:2imp}
\end{eqnarray}

\begin{figure}
\centerline{\epsfxsize=7cm
\epsfbox{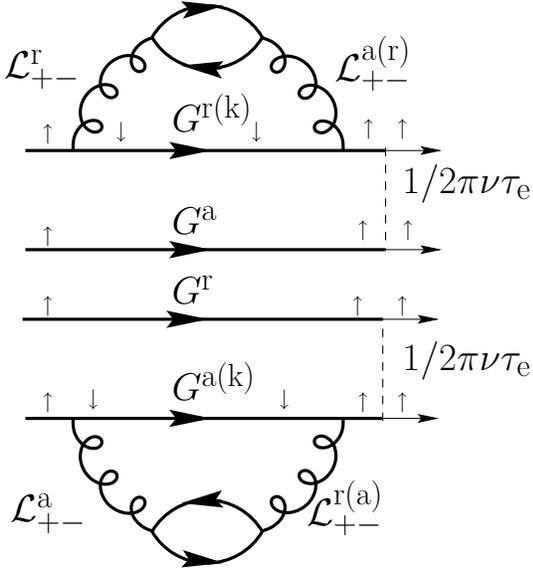}}
\caption {
The Cooperon self energy diagrams to the fourth 
order in the exchange constant $J$. The spin-spin correlation function
(wavy line) acquires correction due to the electron scattering off the spin.
 }
\label{fig:RPA}
\end{figure}

%%%%%%%%%%%%%%%%%%%%%%%%%%%%%%%%%%%%%%%%%%%%%%%%%%%%%%%%%%%%%%%%%%%%%%%%%%%%%%
%%%%%%%%%%%%%%%%%%%%%%%%%%%%%%%%%%%%%%%%%%%%%%%%%%%%%%%%%%%%%%%%%%%%%%%%%%%%%%
%%%%%%%%%%%%%%%%%%%%%%%%%%%%%%%%%%%%%%%%%%%%%%%%%%%%%%%%%%%%%%%%%%%%%%%%%%%%%%
%%%%%%%%%%%%%%%%%%%%%%%%%%%%%%%%%%%%%%%%%%%%%%%%%%%%%%%%%%%%%%%%%%%%%%%%%%%%%%
%%%%%%%%%%%%%%%%%%%%%%%%%%%%%%%%%%%%%%%%%%%%%%%%%%%%%%%%%%%%%%%%%%%%%%%%%%%%%%
%%%%%%%%%%%%%%%%%%%%%%%%%%%%%%%%%%%%%%%%%%%%%%%%%%%%%%%%%%%%%%%%%%%%%%%%%%%%%%
%%%%%%%%%%%%%%%%%%%%%%%%%%%%%%%%%%%%%%%%%%%%%%%%%%%%%%%%%%%%%%%%%%%%%%%%%%%%%%
%%%%%%%%%%%%%%%%%%%%%%%%%%%%%%%%%%%%%%%%%%%%%%%%%%%%%%%%%%%%%%%%%%%%%%%%%%%%%%

\section{Conductance fluctuations}
\label{sec:5}

%%%%%%%%%%%%%%%%%%%%%%%%%%%%%%%%%%%%%%%%%%%%%%%%%%%%%%%%%%%%%%%%%%%%%%%%%%%%%%
%%%%%%%%%%%%%%%%%%%%%%%%%%%%%%%%%%%%%%%%%%%%%%%%%%%%%%%%%%%%%%%%%%%%%%%%%%%%%%
%%%%%%%%%%%%%%%%%%%%%%%%%%%%%%%%%%%%%%%%%%%%%%%%%%%%%%%%%%%%%%%%%%%%%%%%%%%%%%
%%%%%%%%%%%%%%%%%%%%%%%%%%%%%%%%%%%%%%%%%%%%%%%%%%%%%%%%%%%%%%%%%%%%%%%%%%%%%%
%%%%%%%%%%%%%%%%%%%%%%%%%%%%%%%%%%%%%%%%%%%%%%%%%%%%%%%%%%%%%%%%%%%%%%%%%%%%%%
%%%%%%%%%%%%%%%%%%%%%%%%%%%%%%%%%%%%%%%%%%%%%%%%%%%%%%%%%%%%%%%%%%%%%%%%%%%%%%
%%%%%%%%%%%%%%%%%%%%%%%%%%%%%%%%%%%%%%%%%%%%%%%%%%%%%%%%%%%%%%%%%%%%%%%%%%%%%%
%%%%%%%%%%%%%%%%%%%%%%%%%%%%%%%%%%%%%%%%%%%%%%%%%%%%%%%%%%%%%%%%%%%%%%%%%%%%%%

In this Section we study the effect of the impurity spin polarization on conductance fluctuations of a metal ring,
weakly connected to the leads.\cite{AS} 
Magnetic flux $\Phi$ piercing the ring
changes wave functions of electrons in the ring and, consequently, the ring conductance. The fluctuations of the conductance are usually characterized by the correlation function ${\cal K}(\Dt,\Df)$:
\begin{equation}
{\cal K}(\Dt,\Df)=
\langle\langle
g_{\Phi}(0)g_{\Phi+\Delta \Phi}(\Dt)
\rangle\rangle_{\Phi},
\label{eq:30}
\end{equation}
where $\Phi=\pi R^2 H$ is the magnetic flux through the ring of radius $R$, and $\Dt$ is the time lapse  between the measurements of  $g_{\Phi}$ and $g_{\Phi+\Delta \Phi}$. The d.c. conductance, Eq.~(\ref{eq:30}), is defined in terms of the current $I(t)$, averaged over  measurement time $\Tm$
\begin{equation}
g_\Phi(t)=\int_t^{t+\Tm} \left. 
\frac{\partial I_\Phi(t)}{\partial V}\right|_{V=0} d t',
\end{equation}
where $V$ is the applied bias.
During the measurement time $\Tm$ the magnetic
flux should be constant, otherwise the measured conductance is already
averaged over different realizations.

We assume that the system satisfies the ergodic hypothesis and the
averaging over the magnetic field is equivalent to averaging over
impurity configurations.\cite{A1,AKL} To calculate the conductance
correlation function, Eq.~(\ref{eq:30}), we apply the conventional
averaging technique.\cite{AGD} We consider only the case of short
relaxation time $\tau_{\rm T}\ll \Dt$. In this case the spin
correlation functions in Eqs.~(\ref{eq:500}) should be taken in the
limit $\tau\to \infty$, so that $\chi_\bot(\tau\to \infty)=0$ and
$\chi_z(\tau\to \infty)=\Sodin ^2$.

\begin{figure*}
\centerline{
\epsfxsize=10cm
\epsfbox{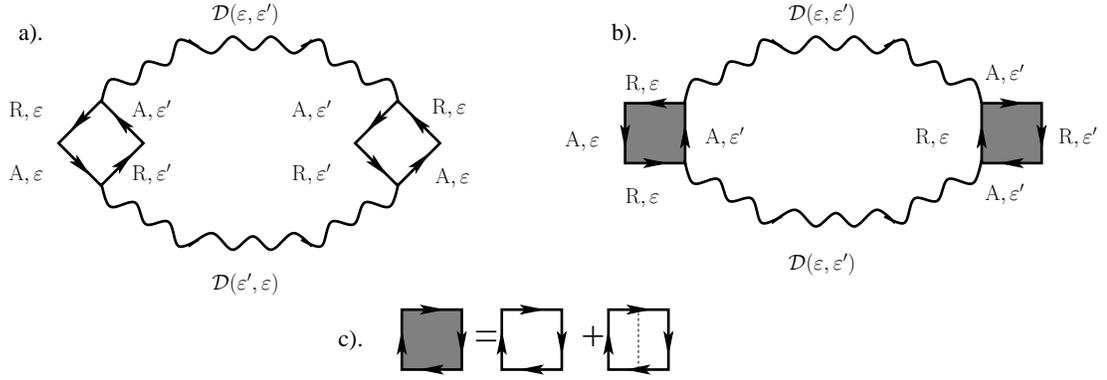}}
\caption {
  Diagrams a) and b) give the main contribution to the conductance
  correlation function. Diagram a) originates from the fluctuations of
  the diffusion coefficient and decreases only as $1/T$ at high
  temperature.  Diagram b) is referred to as the contribution from the
  fluctuations of the electron density of states and exponentially
  vanishes at high temperature, see \cite{AS}.  The Hikami box (black
  square) is the sum of two diagrams, shown in Fig. c).  }
\label{fig:CF-diag}
\end{figure*}

The conductance correlation function contains all harmonics
\begin{equation}
{\cal K}(\Df) = \sum_{k=0}^{\infty}{\cal K}_k
\cos\left(2\pi k \frac{\Df}{\f0}\right).
\label{eq:CF-Hcs}
\end{equation}
The amplitudes ${\cal K}_k$ consist of two parts,\cite{AS} originating from the
fluctuations of the diffusion coefficient, 
Fig.~\ref{fig:CF-diag}a, and the fluctuations of the electron density
of states, Fig.~\ref{fig:CF-diag}b:
\begin{eqnarray}
{\cal K}_k & = & \alpha \frac{e^4}{(2\pi \hbar)^2}
\frac{D^2}{R^4}
\!\! \int\frac{d\e d\e'}{16T^2}\cosh^{-2}\frac{\e}{2T}
\cosh^{-2}\frac{\e'}{2T}
\nonumber
\\
& \times &\sum\limits_{i=1}^4
\left(
{\cal A}^{(i)}_k(\e,\e')+{\cal B}^{ (i)}_k(\e,\e')
\right),
\label{eq:31:D}
\end{eqnarray}
see Appendix~\ref{app:D}. Here $\alpha$ is a geometry dependent
dimensionless factor of order of unity; functions 
${\cal A}^{(i)}_k(\e,\e')$ and ${\cal B}^{(i)}_k(\e,\e')$ are defined by 
\begin{subequations}
\label{eq:MCF-R}
\begin{eqnarray}
\!\!\!\!\!\!\!\!{\cal A}_k^{(i)}(\e,\e')\!\!&\!\! =\!\! &\!\! 2 \!\!\int\!\! 
\left| {\cal D}_i\left(\e,\e',\frac{x}{R}\right) \right|^2
\! \cos(2\pi k x) dx ,
\label{eq:MCF-Ra}
\\
\!\!\!\!\!\!\!\!{\cal B}_k^{({i})}(\e,\e')\!\!&\!\! =\!\! &\!\!
 \!\!\int\!\! 
{\rm Re} \left\{ {\cal D}^2_i\left(\e,\e',\frac{x}{R}\right)\right\}
\cos (2\pi k x)dx 
\label{eq:MCF-Rb}
\end{eqnarray}
\end{subequations}
in terms of the diffuson components:
\begin{equation}
\!\! {\cal D}_i(\e,\e',q)=
\frac{1}{i(\e\! -\! \e'\! +\! \varsigma_{i}\e_{\rm Z})\!+\! Dq^2\! +\! \gd\! +\! \Gamma_i(\e,\e')}.
\label{eq:diff}
\end{equation}

The index $i$ in Eqs.~(\ref{eq:MCF-R}) runs over different spin
configurations of the diffuson ${\cal D}_i(\e,\e',q)$, 
which are related to the
common classification of spin wave functions of two spin-$1/2$
particles in terms of triplet and singlet states, see Table
\ref{tab:2}. In our notations, modes $i=1,2$ correspond to total spin
$S_{\rm d}=1$ with non-zero projections on the magnetic field 
$m_{\rm d}=\pm 1$, mode $i=3$ represents total spin $S_{\rm d}=1$ with zero
projection on the magnetic field, $m_{\rm d}=0$, and mode $i=4$ is a
singlet spin configuration.  Energies $\varsigma_{i}\e_{\rm Z}$
represent the effect of the Zeeman splitting of conduction electron
states on various diffuson modes.  Coefficients $\varsigma_{i}$
are given in Table~\ref{tab:2}.

We explicitly separated two additive components, $\Gamma_i(\e,\e')$
and $\gd$, to the diffuson decay rate in Eq.~(\ref{eq:diff}). The component
$\Gamma_i(\e,\e')$ corresponds to the contributions from spin-orbit
interaction and scattering off magnetic impurities. The component
$\gd$ takes into account other processes, such as electron escape
through the leads and scattering caused by electron-phonon and
electron-electron interactions.\cite{AB} 

\begin{table}
\caption{The decay rates of various components of the diffuson in a metal without spin orbit interaction. 
The components with $i=1,2,3$ represent a spin states with total spin one,
$S_{\rm d}=1$. These spin configurations form a triplet in an
isotropic system. Anisotropy due to the applied magnetic field lifts
the degeneracy. The $i=4$ component is a singlet spin configuration, 
$S_{\rm d}=0$. The last column shows
coefficients $\varsigma_{i}\ $ for the Zeeman splitting of
conduction electrons.} 
\begin{ruledtabular}
\begin{tabular}{|c|ccc|}
$\ i\ $ & $|S_{\rm d},m_{\rm d}\rangle $ 
    & $\Gamma_{i}(H)\ts$ & $\varsigma_{i}\ $ \\
\hline 
$\ 1\ $ & $S_{\rm d}=1, m_{\rm d}=+ 1$ 
& $\displaystyle 1+\Sodin ^2/S^2$ & $ +1\ $ \\ 
$  2  $ & $S_{\rm d}=1, m_{\rm d}=- 1$ 
& $1+\Sodin ^2/S^2$ & $ -1\ $ \\ 
$  3  $ & $S_{\rm d}=1, m_{\rm d}=\mbox{ }0$   
& $ 1-\Sodin ^2/S^2     $ & $ 0 $ \\ 
$  4  $ & $S_{\rm d}=0, m_{\rm d}=\mbox{ }0$   
& $1-\Sodin ^2/S^2  $ & $ 0 $ \\
\end{tabular}
\end{ruledtabular}
\label{tab:2}
\end{table}

%%%%%%%%%%%%%%%%%%%%%%%%%%%%%%%%%%%%%%%%%%%%%%%%%%%%%%%%%%%%%%%%%%%%%%%%%%%%%%
%%%%%%%%%%%%%%%%%%%%%%%%%%%%%%%%%%%%%%%%%%%%%%%%%%%%%%%%%%%%%%%%%%%%%%%%%%%%%%
%%%%%%%%%%%%%%%%%%%%%%%%%%%%%%%%%%%%%%%%%%%%%%%%%%%%%%%%%%%%%%%%%%%%%%%%%%%%%%
\subsection{Effect of classical spin impurities}
\label{sec:5.1}
%%%%%%%%%%%%%%%%%%%%%%%%%%%%%%%%%%%%%%%%%%%%%%%%%%%%%%%%%%%%%%%%%%%%%%%%%%%%%%
%%%%%%%%%%%%%%%%%%%%%%%%%%%%%%%%%%%%%%%%%%%%%%%%%%%%%%%%%%%%%%%%%%%%%%%%%%%%%%
%%%%%%%%%%%%%%%%%%%%%%%%%%%%%%%%%%%%%%%%%%%%%%%%%%%%%%%%%%%%%%%%%%%%%%%%%%%%%%

Following the spirit of Section~\ref{sec:4}, we first analyze
conductance fluctuations in the presence of classical $S\gg 1$ spins.
This analysis allows us to explore the effect of the impurity spin
polarization at various relations between parameters of the system. In
the next subsection we calculate the amplitude of conductance
correlation function, Eqs.~(\ref{eq:MCF-R}), for spins $S\sim 1$ in a
metal with strong spin orbit interaction.

We consider systems with fast spin relaxation time $\tau_{\rm T}\ll
\Dt$, where $\Dt$ is the time lapse between the current measurements,
see Eq.~(\ref{eq:30}). In this case the solution of the diffuson equation,
Eq.~(\ref{eq:10}), is described by the following energy-independent
decay rates:
\begin{subequations}
\label{eq:41}
\begin{eqnarray}
\Gamma_{1,2} & = &  
\left(1+\Sodin ^2/S^2\right)\frac{1}{\ts}
+\frac{4}{3}\gso,
\label{eq:41a}
\\
\Gamma_{3} & = & 
\left(1-\Sodin ^2/S^2\right)\frac{1}{\ts}+\frac{4}{3}\gso,
\label{eq:41b}
\\
\Gamma_{4} & = &\left(1-\Sodin ^2/S^2\right)\frac{1}{\ts}.
\label{eq:41c}
\end{eqnarray} 
\end{subequations}
Here $\Sodin $ is defined by Eq.~(\ref{eq:6a}) and depends only on the
ratio of the magnetic field  and temperature $B/T$.  

The $i=1,2$ modes represent the interference of electron states with opposite spin orientations. The exchange field of magnetic impurities produces different (opposite) contributions to the phases of these two states. Because the phase contributions for different electron trajectories fluctuate, the interference of electron states with opposite spins is suppressed even if all spins of magnetic impurities are fully polarized. 
Thus, the polarization of impurity spins does not suppress the effect of electron phase relaxation due to the scattering off magnetic impurities for diffuson modes $i=1,2$. According to Eq.~(\ref{eq:41a}), the diffuson decay rates $\Gamma_{1,2}$ actually increase as the applied magnetic field increases. 

On the other hand, the $i=3,4$ modes stand for the interference of two electron states with parallel spins. At strong magnetic field 
$B\gg T/g\muB S$, when all spins are polarized, the scattering off magnetic impurities  provides equal phase shifts to both states and does not affect diffuson relaxation rates $\Gamma_{3,4}$. That is why the contribution to $\Gamma_{3,4}$, caused by the scattering off magnetic impurities,  vanishes as impurity spins become polarized, see Eqs.~(\ref{eq:41b}) and (\ref{eq:41c}).

Substituting the diffuson decay times from Eqs.~(\ref{eq:41}) into
Eqs.~(\ref{eq:31:D}) -- (\ref{eq:diff}), we can describe the harmonics of the conductance correlation function Eq.~(\ref{eq:CF-Hcs}) at an arbitrary value of magnetic field, ranging from $B=0$ to $B\gg T/g\muB S$. To analyze this crossover, we first consider a metal without spin-orbit interaction, $\gso=0$, and evaluate the integrals over energies $\e$ and $\e'$ in
Eq.~(\ref{eq:31:D}) for two limiting cases of low ($T\ll
\Gamma_i+\gd$) or high temperature ($T\gg \Gamma_i+\gd$).  

In the low temperature limit we have:\cite{AS}
\begin{eqnarray}
{\cal K}_k  = 
\frac{3\pi\alpha}{2}\frac{e^4}{(2\pi\hbar)^2}  \sum\limits_{i=1}^{4} 
\frac{L_i^3}{R^3}  \left(2\pi k\frac{R}{L_i}+1\right)
e^{-2\pi k R/L_i}
\label{eq:CF-LT}
\end{eqnarray}
with 
\begin{equation}
L_i=\sqrt{\frac{D}{\Gamma_i+\gd}}.
\label{eq:deph-length}
\end{equation}
For simplicity, we omitted  the Zeeman splitting of the conduction
electron states, which actually modifies the $i=1,2$ terms in
Eq.~(\ref{eq:CF-LT}).  
We emphasize that at low temperature both diagrams in Fig.~\ref{fig:CF-diag}a and \ref{fig:CF-diag}b contribute to the conductance correlation
function.  

In the limit of high temperature $T\gg \Gamma_i+\gd$, 
the contribution of the diagram in
Fig.~\ref{fig:CF-diag}b is  small as 
$\exp [-(2\pi)^{3/2} k R(T/D)^{1/2}]$ and may be disregarded. 
The contribution due to fluctuations of the diffusion coefficient,
diagram in Fig.~\ref{fig:CF-diag}a, decays only as $1/T$ at high
temperature $T$:
\begin{eqnarray}
{\cal K}_k  =\alpha \frac{\pi^2}{3} \frac{e^4}{(2\pi\hbar)^2}
\sum\limits_{i=1}^4
f\left(\frac{\varsigma_i \e_{\rm Z}}{2T}\right)
\frac{L_T^2L_i}{R^3} e^{-2\pi k R/L_i},
\label{eq:39}
\end{eqnarray}
where $L_T=\sqrt{D/T}$ is the thermal length, and coefficients $\varsigma_i$
are presented in Table~\ref{tab:2}. The function
\begin{equation}
f\left(z\right)=3\frac{z\cosh z-\sinh z}{\sinh^3 z}
\end{equation} 
takes into account the Zeeman splitting; $f(0)=1$ and 
$f(z)\approx 12(z-1)e^{-2z}$ for $z\gg 1$.

Regardless of scattering off magnetic impurities,  the Zeeman splitting  destroys the contribution of modes $i=1,2$ to the conductance correlation function ${\cal K}$ at $\e_{\rm Z}\gg T$, see Eq.~(\ref{eq:39}). 
From Eq.~(\ref{eq:39}) we conclude that at strong magnetic field, 
$\e_{\rm Z}\gg T$ and $S\os\gg T$,  the amplitude of conductance fluctuations is no longer determined by the scattering rate off magnetic impurities, $1/\ts$. Since modes $i=1,2$ are also suppressed at $\e_{\rm Z}\gg T$
even in a metal without spin impurities, we conclude that 
a strong magnetic field restores the amplitude of conductance fluctuations up to the amplitude in the same ring as if magnetic impurities were absent.

We also discuss another effect of the Zeeman splitting on the conductance correlation function. If the exchange constant $J$ is antiferromagnetic, $J>0$, the Zeeman splitting vanishes not only at $B=0$ but also at some finite $B$, see Eq.~(\ref{zeeman}), and thus produces another wrinkle in the
dependence of ${ \cal K}_k$ on the applied magnetic field. 
This non-monotonic behavior of the amplitude of
the conductance oscillations is reminiscent to the reentrance effect
in superconductors.\cite{Jaccarino}

\begin{figure}
\centerline{\epsfxsize=7.5cm
\epsfbox{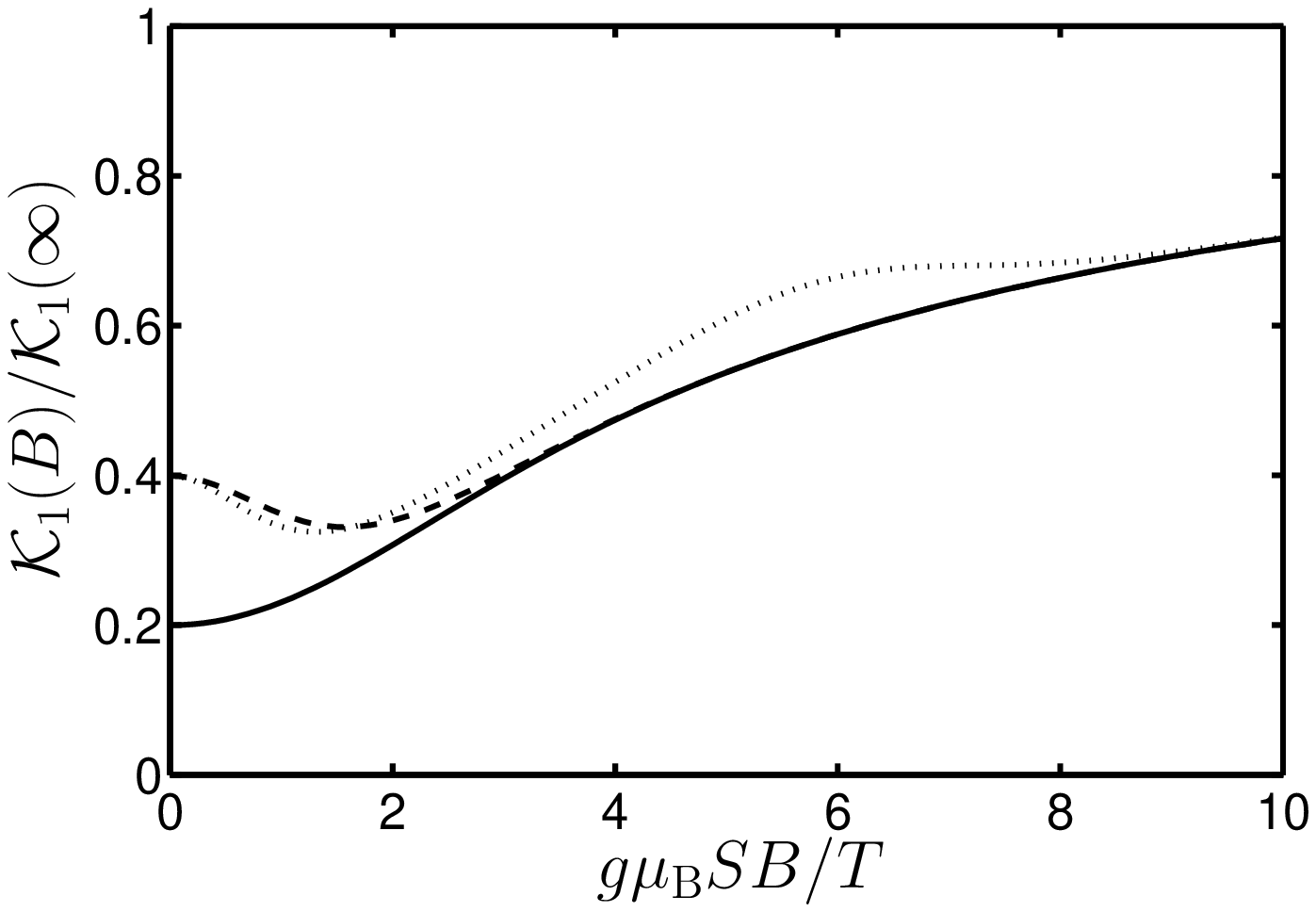}}
\caption{Amplitude ${\cal K}_1(B)$  as a
  function of the applied magnetic field $B$ for a metal with
  classical ($S\gg 1$) magnetic impurities in the absence of spin
  orbit interaction. A non-monotonic behavior of ${\cal K}_1$ due to
  the Zeeman splitting of the conduction electron states is
  illustrated for two cases: \emph{i)} $\e_{\rm Z}=\os$ (dashed line),
  \emph{ii)} $\e_{\rm Z}=\os-7T\Sodin$ (dotted line), here the second
  term represents the average exchange field $2n_{\rm s}J\Sodin$ with
  $n_{\rm s}J=3.5T$. The limit $\e_{\rm Z}\to\infty$ is presented by a
  solid line. We choose $\gd\ts=2$ and $R=\sqrt{D/\gd}$.}
\label{fig:CF-cl}
\end{figure}

To illustrate the effects of the Zeeman splitting $\e_{\rm Z}$, we plot
amplitude ${\cal K}_1$ of the principal harmonic, $k=1$, as a function
of $B$ in Fig.~\ref{fig:CF-cl}. We choose the following values of the
system parameters: $\gd\ts=2$ and $R=\sqrt{D/\gd}$.  
The dashed line corresponds to the case of equal g-factors for the
conduction electrons and impurity spins, {\it i.e.},
$\varepsilon_{\rm Z}=\os$, and negligibly small exchange contribution
in Eq.~(\ref{zeeman}), $n_sJ\langle S\rangle\ll\omega_S$. The limit of
$\e_{\rm Z}\gg T$, when only terms $i=3,4$ in Eq.~(\ref{eq:39})
survive, is represented by the solid line. Finally, the dotted line
demonstrates the reentrance effect due to the antiferromagnetic
impurities at some specific value of $n_{\rm s}J =3.5 T$ with $T$
being the temperature.

Now we notice, see Eqs.~(\ref{eq:41}), that the spin-orbit interaction suppresses the
contribution to the amplitude of the conductance fluctuations,
originating from the diffuson modes with $i=1,2,3$ ($S_{\rm d}=1$). 
In the limit of strong spin-orbit interaction ($\gso\gg D/R^2$) only
the contribution from the singlet ($S_{\rm d}=0$ and $i=4$) mode
survives:
\begin{eqnarray}
{\cal K}_k & = &\alpha \frac{\pi^2}{3}\frac{e^4}{(2\pi\hbar)^2}
\frac{L_T^2L_4}{R^3} e^{-2\pi k R/L_4},
\label{eq:39SO}
\end{eqnarray} 
with length $L_4$ defined by Eqs.~(\ref{eq:41c}) and (\ref{eq:deph-length}). 
The function ${\cal K}_k$ for a metal with strong spin
orbit scattering monotonically increases as a function of $B/T$,
provided $\gd$ remains constant. At strong magnetic field, $S\os\gg
T$, magnetic impurities do not affect the conductance fluctuations.

We emphasize that the non-monotonic behavior of the amplitudes 
${\cal K}_k$ of the conductance correlation function originates from the
diffuson modes with $i=1,2$, which do not contribute to the
correlation function in metals with strong spin-orbit interaction.
Thus, the monotonic or non-monotonic behavior of the amplitudes 
${\cal K}_k$ distinguishes materials with or without spin orbit coupling.

%%%%%%%%%%%%%%%%%%%%%%%%%%%%%%%%%%%%%%%%%%%%%%%%%%%%%%%%%%%%%%%%%%%%%%%%%%%%%%
%%%%%%%%%%%%%%%%%%%%%%%%%%%%%%%%%%%%%%%%%%%%%%%%%%%%%%%%%%%%%%%%%%%%%%%%%%%%%%
%%%%%%%%%%%%%%%%%%%%%%%%%%%%%%%%%%%%%%%%%%%%%%%%%%%%%%%%%%%%%%%%%%%%%%%%%%%%%%
\subsection{Effect of quantum spin impurities}
\label{sec:5.2}
%%%%%%%%%%%%%%%%%%%%%%%%%%%%%%%%%%%%%%%%%%%%%%%%%%%%%%%%%%%%%%%%%%%%%%%%%%%%%%
%%%%%%%%%%%%%%%%%%%%%%%%%%%%%%%%%%%%%%%%%%%%%%%%%%%%%%%%%%%%%%%%%%%%%%%%%%%%%%
%%%%%%%%%%%%%%%%%%%%%%%%%%%%%%%%%%%%%%%%%%%%%%%%%%%%%%%%%%%%%%%%%%%%%%%%%%%%%%
In this section we perform a quantum calculation to analyze the effect of polarization of impurity spins with $S\sim 1$, when the semiclassical
description of electron scattering off magnetic impurities is not
applicable. We consider metals with strong spin orbit
interaction, $\gso\gg D/R^2$, so that only the singlet component of
the diffuson survives.  In this case the calculation of the diffuson
self energy is similar to the calculation of the proper quantity for
the Cooperon, see Sec.\ref{sec:4.3}. Details of calculations and the
expression for the amplitude of the conductance fluctuations are
presented in Appendix \ref{app:E}.

\begin{figure}
\centerline{\epsfxsize=7.5cm
\epsfbox{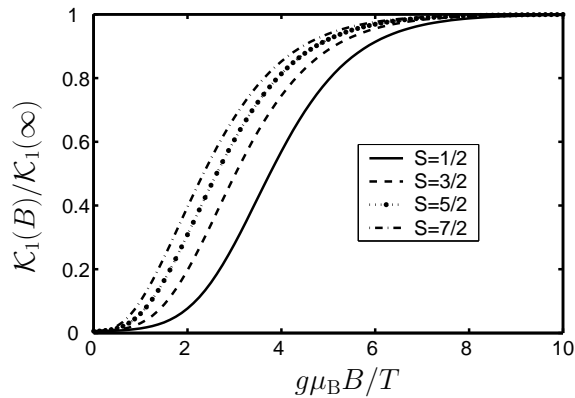}}
\caption{ The dependence of conductance oscillations ${\cal K}_1(B)$ as a function of the applied magnetic field $B$ for four values of impurity spins:
$S=1/2$ (solid line), $S=3/2$ (dashed line), $S=5/2$ (dotted line), $S=7/2$ (dash-dotted line). We choose $\gd\ts=1/2$ and $R=\sqrt{D/\gd}$.}
\label{fig:CF}
\end{figure}

Below we focus on the high temperature case,  $T\gg \Gamma(\e,\e'\simeq T)$, which is reached at sufficiently strong magnetic field even if $\ts T\ll 1$.
At high temperature, the integral over the difference of $\e$ and $\e'$
converges fast, $|\e-\e'|\lesssim \Gamma(\e,\e'\simeq T)$. This observation
allows us to perform integration over the energy difference. We obtain:
\begin{eqnarray}
{\cal K}_k & = & \alpha \frac{\pi^2}{8}\frac{e^4}{(2\pi\hbar)^2}\frac{\sqrt{D}L_T^2}{R^3}
\nonumber
\\
&& \times
\int
\frac{e^{-2\pi kR\sqrt{\Gamma(\e)+\gd}/\sqrt{D}}}{\sqrt{\Gamma(\e)+\gd}}
\frac{d\e/T}{\cosh^4\e/2T},
\label{11'}
\end{eqnarray}
with the dephasing rate in the form of
\begin{equation}
\Gamma(\e)= \left[ 1 -
\frac{\langle \hat{S}_{z}\rangle^{2}+ \langle \hat{S}_{z}\rangle
\tanh(\e+\omega_{\rm{s}})/2T }{S(S+1)}
\right]\frac{1}{\ts}.
\label{eq:dqdt}
\end{equation}
We represent the corresponding curves of ${\cal K}_1$ as a function of the applied magnetic field $B$ in Fig.~\ref{fig:CF} for several values of $S$, assuming $\ts T\gg 1$ and for the following values of the system parameters: $\gd\ts=1/2$ and $R=\sqrt{D/\gd}$. The shapes of the curves corresponding to various values of the impurity spin $S$ 
are different from each other. Particularly, we conclude that the conductance fluctuations are faster restored by magnetic field in metals with a larger value of  $S$. The fitting of experimental results by the curves ${\cal K}_k(B)$ given by Eq.~(\ref{11'}), may provide the impurity spin parameters, such as its value $S$ and the gyromagnetic factor $g$.

As the polarization $\Sodin$  and the Zeeman splitting $\os$ increase, the diffuson decay rate $\Gamma(\e\simeq T)$ to the second order in the exchange constant $J$ becomes exponentially small, see Eq.~(\ref{eq:dqdt}). In this case we have to take into account higher-order contribution in $J$, which is represented by Eq.~(\ref{eq:SE2}). The crossover from the phase relaxation rate in the form of Eq.~(\ref{eq:dqdt}) to the form of Eq.~(\ref{eq:SE2}) takes place at magnetic field $B\sim B_*(T)$, see Eq.~(\ref{eq:1.7}).

%%%%%%%%%%%%%%%%%%%%%%%%%%%%%%%%%%%%%%%%%%%%%%%%%%%%%%%%%%%%%%%%%%%%
%%%%%%%%%%%%%%%%%%%%%%%%%%%%%%%%%%%%%%%%%%%%%%%%%%%%%%%%%%%%%%%%%%%%
%%%%%%%%%%%%%%%%%%%%%%%%%%%%%%%%%%%%%%%%%%%%%%%%%%%%%%%%%%%%%%%%%%%%
\section{Conclusions}
\label{sec:6}
%%%%%%%%%%%%%%%%%%%%%%%%%%%%%%%%%%%%%%%%%%%%%%%%%%%%%%%%%%%%%%%%%%%%
%%%%%%%%%%%%%%%%%%%%%%%%%%%%%%%%%%%%%%%%%%%%%%%%%%%%%%%%%%%%%%%%%%%%
%%%%%%%%%%%%%%%%%%%%%%%%%%%%%%%%%%%%%%%%%%%%%%%%%%%%%%%%%%%%%%%%%%%%
The main question addressed in this paper, is how the polarization of magnetic impurities in a metal affects the interference contribution to its conductance. It is well-known that at weak magnetic field, which does not cause such polarization, the electron scattering off localized spins results in suppression of the weak localization correction to the conductivity and in suppression of the mesoscopic conductance fluctuations. If these interference contributions can be restored by an application of a stronger spin-polarizing field, then a transport measurement may serve as a test for the presence of magnetic impurities in a sample. This possibility was the main motivation of the presented study. 

We obtained analytical results for the WL correction to the conductivity and for the amplitude of conductance fluctuations valid for an arbitrary magnetic field. We found that the conditions for the restoration of weak localization correction are quite stringent. The weak localization correction can be substantially enhanced by an application of a magnetic field only in samples of very small size (to avoid the orbital effect of the magnetic field) and made of a light-element material, to make the spin-orbit scattering negligible. If these conditions are met, the spin polarization may completely eliminate the effect of magnetic impurities. At intermediate fields $B$, the weak localization correction depends on the ratio $B/T$, and we find the corresponding crossover function spanning the full range of spin polarizations. 

The effect of spin polarization on the mesoscopic conductance fluctuations is much more robust. The amplitude of fluctuations is restored if variations of a spin configuration in time are suppressed by the external magnetic field. At the same time, there is no detrimental orbital influence of the applied magnetic field on mesoscopic fluctuations. The spin-orbit interaction also does not spoil the effect of spin polarization on the amplitude of fluctuations. We evaluated the amplitude of the ``$h/e$'' Aharonov-Bohm  oscillations of the conductance correlation function.  In a metal with strong spin orbit interaction, the polarization of magnetic impurities restores this amplitude up to its value characteristic for the host material in the absence of magnetic impurities. We find the full crossover function, which describes the amplitude of fluctuations at arbitrary value of $B/T$. The use of our results for the analysis of transport measurements may, in principle, yield such characteristics of magnetic impurities as the values of their spin  and  $g$-factor, which are hard to access directly.

\section*{Acknowledgements}
We acknowledge discussions with M. Reznikov, which initiated the present work. We are also grateful to I. Aleiner, N.  Birge, V. Falko, A. Kamenev, J. Meyer, P. Mohanty and H. Pothier  for valuable discussions.  
We thank F. Pierre and N. Birge for providing us with their preprint\cite{Birge} before its publication.
This work was supported by NSF Grants No. DMR
97-31756, DMR 012070 and EIA 0210736.

\appendix
%%%%%%%%%%%%%%%%%%%%%%%%%%%%%%%%%%%%%%%%%%%%%%%%%%%%%%%%%%%%%%%%%%%%
%%%%%%%%%%%%%%%%%%%%%%%%%%%%%%%%%%%%%%%%%%%%%%%%%%%%%%%%%%%%%%%%%%%%
%%%%%%%%%%%%%%%%%%%%%%%%%%%%%%%%%%%%%%%%%%%%%%%%%%%%%%%%%%%%%%%%%%%%
\section{Cooperon and diffuson in a metal with classical magnetic impurities}
\label{app:A}
%%%%%%%%%%%%%%%%%%%%%%%%%%%%%%%%%%%%%%%%%%%%%%%%%%%%%%%%%%%%%%%%%%%%
%%%%%%%%%%%%%%%%%%%%%%%%%%%%%%%%%%%%%%%%%%%%%%%%%%%%%%%%%%%%%%%%%%%%
%%%%%%%%%%%%%%%%%%%%%%%%%%%%%%%%%%%%%%%%%%%%%%%%%%%%%%%%%%%%%%%%%%%%
In calculations of the weak localization correction to the conductivity of
a metal with classical magnetic moments, we use the Cooperon which is
formally defined as an average of electron Green's functions
\begin{widetext}
\begin{eqnarray}
{\cal C}^{\alpha\beta}_{\gamma\delta}
\left(\frac{t_1^+-t_1^-}{2},\frac{t_2^+-t_2^-}{2},
t_1^++t_1^-;\r,\r'\right)
\delta(t_1^++t_1^--t_2^+-t_2^-)
& = &
\langle
G^{(\rm R)}_{\alpha\beta}(t_1^+,t_2^+;\r,\r') 
G^{(\rm A)}_{\gamma\delta}(t_1^-,t_2^-;\r,\r')\rangle.
\label{eq:8}
\end{eqnarray}
After the standard procedure\cite{WL} we obtain the following
expression for the Cooperon:
\begin{subequations}
\label{eq:9}
\begin{eqnarray}
\left[\left(\frac{\partial}{\partial \tau}+D\q^2\right)\hat 1 + 
\hat {\cal H}_{0}^{\cal C}
+\hat {\cal H}_{\rm s}^{\cal C}(\tau)\right]
\hat {\cal C}(\tau,\tau',T;\q)=\hat 1  \delta(\tau-\tau'),
\label{eq:9a}
\end{eqnarray}
where $\hat {\cal H}^{\cal C}_0$  is a matrix in spin space 
representing the Zeeman splitting of conduction electron states and spin-orbit interaction:
\begin{equation}
\label{eq:9b}
\hat {\cal H}_{0}^{\cal C}=
\left(
\begin{array}{cccc}
 0 & 0 & 0 & 0 \\ 0 & 2i\e_{\rm Z} & 0 & 0 \\ 0 & 0 & -2i\e_{\rm Z} & 0 \\
 0 & 0 & 0 & 0
\end{array}
\right)
+
\frac{2\gso}{3} \left(
\begin{array}{cccc}
2 & 0 & 0 & 0 \\ 0 &1 & 1 & 0 \\ 0 & 1 & 1 & 0 \\ 0 & 0 & 0 &
2
\end{array}
\right),
\end{equation}
and $\hat {\cal H}^{\cal C}_{rm s}$ describes scattering off magnetic impurities:
\begin{equation}
\label{eq:9c}
\hat {\cal H}_{\rm s}^{\cal C}(\tau)=\frac{1}{\ts}
\left(
\begin{array}{cccc}
 (1-\chi_{zz}(\tau)) & 0 & 0 & 0 \\ 0 & (1+\chi_{zz}(\tau))) &
 -\chi_{\bot}(\tau)) & 0 \\ 0 & -\chi^*_{\bot}(\tau)) &
 (1+\chi_{zz}(\tau))) & 0 \\ 0 & 0 & 0 & (1-\chi_{zz}(\tau)))
\end{array}
\right)
\end{equation}
\end{subequations}
The solution of Eqs.~\ref{eq:9} yields the Cooperon in the form of Eq.~(\ref{eq:11}). 

The diffuson appears in this paper in calculations of the conductance correlation function and is defined as 
\begin{eqnarray}
\label{eq:7}
{\cal D}_{\gamma\delta}^{\alpha\beta}
\left(\frac{t_1^++t_1^-}{2},\frac{t_2^++t_2^-}{2},t_1^+-t_1^-;\r,\r'\right)
\delta(t_1^+-t_1^--t_2^++t_2^-)
& = &
\langle
G^{(\rm R)}_{\alpha\beta}(t_1^+,t_2^+;\r,\r') 
G^{(\rm A)}_{\delta\gamma}(t_2^-,t_1^-;\r',\r)
\rangle.
\end{eqnarray}
It satisfies the following equation:
\begin{subequations}
\label{eq:10}
\begin{eqnarray}
\label{eq:10a}
\left[\left(i\o+D\q^2\right)\hat 1 + \hat {\cal H}_{0}^{\cal D}
+\hat {\cal H}_{\rm s}^{\cal D}(\tau)\right]
\hat {\cal D}_{\q}(\o,\tau)=\hat 1,
\end{eqnarray}
where the matrix $\hat {\cal H}^{\cal D}_0$ does not depend on time variable $\tau$ and describes the Zeeman splitting of conduction electron states and spin-orbit interaction:
\begin{equation}
\label{eq:10b}
\hat {\cal H}_{0}^{\cal D}
\left(
\begin{array}{cccc}
 0 & 0 & 0 & 0 \\ 0 & 2i\e_{\rm Z} & 0 & 0 \\ 0 & 0 & -2i\e_{\rm Z} & 0 \\
 0 & 0 & 0 & 0
\end{array}
\right)
+
\frac{2\gso}{3} \left(
\begin{array}{cccc}
1 & 0 & 0 & 1 \\ 0 & 2 &0 & 0 \\ 0 & 0 & 2 & 0 \\ 1 & 0 & 0 & 1
\end{array}
\right), 
\end{equation}
and $\hat {\cal H}_{\rm s}^{\cal D}(\tau)$ represents scattering off magnetic impurities:
\begin{equation}
\label{eq:10c}
\hat {\cal H}_{\rm s}^{\cal D}(\tau)=\frac{1}{\ts} 
\left(
\begin{array}{cccc}
 (1-\chi_{zz}(\tau)) & 0 & 0 & -\chi_{\bot}(\tau) \\ 0 &
 (1+\chi_{zz}(\tau)) & 0 & 0 \\ 0 & 0 & (1+\chi_{zz}(\tau)) & 0 \\
 -\chi_{\bot}^*(\tau) & 0 & 0 & (1-\chi_{zz}(\tau))
\end{array}
\right).
\end{equation}
\end{subequations}
\end{widetext}

Solving Eqs.~(\ref{eq:10}), we obtain the diffuson in the form 
of Eq.~(\ref{eq:diff}).

%%%%%%%%%%%%%%%%%%%%%%%%%%%%%%%%%%%%%%%%%%%%%%%%%%%%%%%%%%%%%%%%%%
%%%%%%%%%%%%%%%%%%%%%%%%%%%%%%%%%%%%%%%%%%%%%%%%%%%%%%%%%%%%%%%%%
\section{Cooperon decay rate due to scattering off quantum magnetic impurities}
\label{app:B}
%%%%%%%%%%%%%%%%%%%%%%%%%%%%%%%%%%%%%%%%%%%%%%%%%%%%%%%%%%%%%%%%%
%%%%%%%%%%%%%%%%%%%%%%%%%%%%%%%%%%%%%%%%%%%%%%%%%%%%%%%%%%%%%%%%%
%%%%%%%%%%%%%%%%%%%%%%%%%%%%%%%%%%%%%%%%%%%%%%%%%%%%%%%%%%%%%%%%%
In this Appendix we derive the Cooperon decay rate in a metal with quantum ($S\sim 1$) impurity spins. 
The Hamiltonian of interaction of conduction electrons with magnetic impurities, Eq.~(\ref{eq:HS}), 
has the form:
\begin{equation}
{H}_{\rm m}= J\left(
\hat{S}_{z}\hat{\sigma}_{z}+\hat{S}_{-}\hat{\sigma
}_{+}+\hat{S}_{+}\hat{\sigma}_{-}\right)
\label{eq:Hm}
\end{equation}
where $\hat{\sigma}_{x}$, $\hat{\sigma}_{y}$ and $\hat{\sigma}_{z}$ are
the Pauli matrices and $\hat{\sigma}_{\pm}=\left( \hat{\sigma}_{x}\pm
i\hat{\sigma}_{y}\right)/2$.  We notice that the first term in Eq.~(\ref{eq:Hm}) represent scattering without spin flip, while two other terms correspond to spin flip scattering.

\begin{figure}
\centerline{\epsfxsize=8cm
\epsfbox{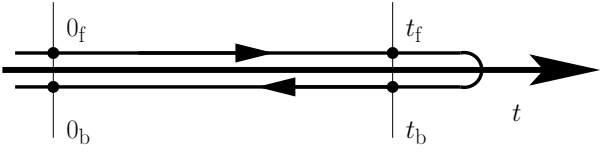}}
\caption {
The Keldysh contour allows to take into account time ordering of quantum
operators. The time subscripts $t_{\rm f,b}$ mean that the operator is
taken at time $t$ at the forward ($t_{\rm f}$) or backward ($t_{\rm b}$)
part of the Keldysh contour.  }
\label{fig:Keldysh_contour}
\end{figure}

We will follow the  standard Keldysh formalism: first, we define electron and
spin operators on the forward and backward parts of the Keldysh contour,
see Fig. \ref{fig:Keldysh_contour}, and then perform rotation in the
Keldysh space. As the result, we obtain the standard form of the electron
Green's function in terms of the retarded
$G^{\mathrm{(R)}}(\varepsilon,{\mathbf{p}})$, advanced
$G^{\mathrm{(A)}}(\varepsilon,{\mathbf{p}})$ and Keldysh
$G^{\mathrm{(K)}}(\varepsilon,{\mathbf{p}})$ components. The spin operator
in the rotated basis acquires the form
\begin{equation}
\hat{\mathcal{S}}_{i}(t)=\frac{1}{2}\left(
\begin{array}
[c]{cc}%
\hat{S}_{i}^{\mathrm{s}}(t) & \hat{S}_{i}^{\mathrm{a}}(t)\\
\hat{S}_{i}^{\mathrm{a}}(t) & \hat{S}_{i}^{\mathrm{s}}(t)
\end{array}
\right)  ,
\end{equation}
where $\hat{S}_{i}^{\mathrm{s,a}}(t)=\hat{S}_{i}(t_{\mathrm{f}})
\pm\hat{S}_{i}(t_{\mathrm{b}})$. In our calculations of the impurity average
conductivity, we work in the Born approximation, therefore the scattering is
completely characterized by the bilinear spin correlators:
\begin{subequations}
\begin{eqnarray}
{\mathcal{L}}_{ij}^{\mathrm{k}}(t) & = & \frac{1}{4}
\langle\hat{S}_{i}^{\mathrm{s}}(t)\hat {S}_{j}^{\mathrm{s}}(t)\rangle,\\
{\mathcal{L}}_{ij}^{\mathrm{r}}(t) & = & \frac{1}{4}
\langle\hat{S}_{i}^{\mathrm{s}}(t)\hat{S}_{j}^{\mathrm{a}}(0)\rangle,\\
{\mathcal{L}}_{ij}^{\mathrm{a}}(t) & = & \frac{1}{4}
\langle\hat{S}_{i}^{\mathrm{a}}(t)\hat {S}_{j}^{\mathrm{s}}(0)\rangle,
\end{eqnarray}
\label{KSF}
\end{subequations}
which are counterparts of the Keldysh Green's function of an
interaction field, see \cite{RS}.

For the Zeeman Hamiltonian
$H_{\rm{spin}}=g\muB B\hat S_{z}$, spin correlator components with
$i=j=z$ have only the Keldysh element
${\mathcal{L}}_{zz}^{\mathrm{k}}(t)$:
\begin{equation}
{\mathcal{L}}_{zz}^{\mathrm{k}}(t)=\langle\hat{S}_{z}^{\mathrm{s}}(t)
\hat{S}_{z}^{\mathrm{s}}(0)\rangle=\langle\hat{S}_{z}^{2}\rangle,
\label{scorr_zz}
\end{equation}
and ${\mathcal{L}}_{zz}^{\mathrm{r,a}}(t)$ vanish due to the commutation
relations of the spin operator along $z$\ axis and the Hamiltonian
$H_{\rm{spin}}$.

Now we present correlators of the spin components perpendicular to the
magnetic field:
\begin{subequations}
\begin{eqnarray}
{\mathcal{L}}_{-+}^{\mathrm{a}}(t) & = &
\theta(-t)e^{i\omega_{\rm{s}} t}\left\langle \hat{S}_{z}\right\rangle ,\\
{\mathcal{L}}_{-+}^{\mathrm{r}}(t) & = & -\theta(t)e^{i\omega_{\rm{s}}
t}\left\langle \hat{S}_{z}\right\rangle ,\\
{\mathcal{L}}_{-+}^{\mathrm{k}}(t) & = & e^{i\omega_{\rm{s}}t}\left(
S(S+1)-\langle\hat{S}_{z}^{2}\rangle\right).
\end{eqnarray}
\label{scorr_pm}
\end{subequations}

Using the formal definitions of Eqs.~(\ref{scorr_zz}) and (\ref{scorr_pm}), we calculate the Cooperon decay rate, which has the meaning of the Cooperon self energy, see Eq.~(\ref{eq:11}). The diagrams which contribute to the decay rates of Cooperon modes $i=1,2$ are shown in Fig.~\ref{fig:Cse}a and \ref{fig:Cse}b and correspond to the electron self-energy due to magnetic
impurities:
\begin{widetext}
\begin{subequations}
\begin{eqnarray}
\Sigma^{\rm r}(\varepsilon) &  = &
\int \frac{d\p}{(2\pi)^d} \int \frac{d\omega}{2\pi}
\left\{\left( {\mathcal{L}} _{-+}^{\rm{k}}(\omega) + 
{\mathcal{L}} _{zz}^{\rm{k}}(\omega)\right)
G^{\rm{r}}(\e+\omega,{\mathbf{p}}) +
\left( {\mathcal{L}} _{-+}^{\rm{r}}(\omega) 
+ {\mathcal{L}} _{zz}^{\rm{r}}(\omega) \right)
G^{\rm{k}}(\e+\omega,{\mathbf{p}})
\right\},
\label{Ese.a}
\\
\Sigma^{\rm a}(\varepsilon) &  = &
\int \frac{d\p}{(2\pi)^d} \int \frac{d\omega}{2\pi}
\left\{\left( {\mathcal{L}} _{-+}^{\rm{k}}(\omega) + 
{\mathcal{L}} _{zz}^{\rm{k}}(\omega)\right)
G^{\rm{a}}(\e+\omega,{\mathbf{p}}) +
\left( {\mathcal{L}} _{-+}^{\rm{a}}(\omega) 
+ {\mathcal{L}} _{zz}^{\rm{a}}(\omega) \right)
G^{\rm{k}}(\e+\omega,{\mathbf{p}})
\right\}.
\label{Ese.b}
\label{eq:Ese}
\end{eqnarray}
\end{subequations}
\end{widetext}\noindent
The elastic part of the self-energy is expressed only in terms of the
retarded and advanced Green's function and is not affected by the
electron distribution, while the inelastic part of the self energy
contains also the Keldysh component of the electron Green's function.
Consequently, for inelastic processes the electron distribution is
important, since such processes should satisfy the fermion exclusion
principle. The contribution to the Cooperon decay rate from diagrams
in Figs.~\ref{fig:CF-diag}a and \ref{fig:CF-diag}b is
$\Gamma_{\rm{ab}}(\varepsilon)=i n_{\rm s} J^{2}
[\Sigma^{\mathrm{a}}(\varepsilon)- \Sigma^{\mathrm{b}}(\varepsilon)]$
and can be written as
\begin{equation}
\Gamma_{\rm{ab}}(\varepsilon)\!=\!\frac{1}{\ts}
\left[\! 1\!
-\!\!\int\! 
\tanh\frac{\varepsilon+\omega}{2T}
\frac{{\rm Im}
\left\{{\cal L}_{-+}^{\rm{r}}(\omega)\right\}}
{S(S+1)}\frac{d\omega }{2\pi}
\right] .
\label{eq:cse_se} 
\end{equation}
The first term in Eq.~(\ref{eq:cse_se}) originates from terms in
Eqs.~(\ref{eq:Ese}), containing the Keldysh components of the spin
correlators ${\mathcal{L}} _{zz}^{\rm{k}}(\omega)$ and ${\mathcal{L}}
_{-+}^{\rm{k}}(\omega)$. We used the  property of these
correlators ${\mathcal{L}} _{zz}^{\rm{k}}(t=0)+{\mathcal{L}}
_{-+}^{\rm{k}}(t=0)=S(S+1)$, arising from their definitions, see
Eqs.~(\ref{KSF}). We observe that the first term in Eq.~(\ref{eq:cse_se})
is the total electron scattering rate on magnetic impurities
$1/\tau_{\rm s}$.  The second term represents the rate of inelastic
scattering processes, accompanied by the energy transfer. Particularly,
this term explicitly contains the electron distribution function in the
form $1-2n(\e)=\tanh\e/2T$, which takes into account the Pauli exclusion
principle for electrons.

We identify the diagram in Fig.~\ref{fig:Cse}c  with the vertex correction
to the Cooperon self energy. For the Cooperon component with parallel
electron spins, only the component of the scattering on magnetic
impurities without spin flip remains:
\begin{eqnarray}
\Gamma_{\rm{c}}(\varepsilon)  
& = & \frac{1}{\ts}
\frac{\langle \hat{S}_{z}^{2}\rangle}{S(S+1)} .
\end{eqnarray}
The decay rate of Cooperon modes $i=1,2$ is given by the sum of
$\Gamma_{\rm{ab}}(\e)$ and $\Gamma_{\rm{c}}(\e)$:
\begin{equation}
\Gamma(\varepsilon)=\frac{1}{\ts}
\left[ 1 -
\frac{\langle \hat{S}_{z}^{2}\rangle
+\langle \hat{S}_{z}\rangle \tanh(\varepsilon+\omega_{\rm{s}})/2T 
}{S(S+1)}
\right].
\label{eq:cse}
\end{equation}
When deriving Eq.~(\ref{eq:cse}) we assumed that the conduction
electron spins and average impurity spins are parallel, so that
$\omega_{\rm{s}}>0$ and $\langle \hat{S}_{z}\rangle >0$. The opposite
case is also described by Eq.~(\ref{eq:cse}) with $\omega_{\rm{s}}<0$
and $\langle \hat{S}_{z}\rangle <0$.  Equation~(\ref{eq:cse})
determines the energy-dependent phase relaxation rate in the WL
correction to conductivity, see Eq.~(\ref{eq:22}).

%%%%%%%%%%%%%%%%%%%%%%%%%%%%%%%%%%%%%%%%%%%%%%%%%%%%%%%%%%%%%%%%%%%%
%%%%%%%%%%%%%%%%%%%%%%%%%%%%%%%%%%%%%%%%%%%%%%%%%%%%%%%%%%%%%%%%%%%%
%%%%%%%%%%%%%%%%%%%%%%%%%%%%%%%%%%%%%%%%%%%%%%%%%%%%%%%%%%%%%%%%%%%%
\section{The $J^4$-order contribution to the electron self energy }
\label{app:C}
%%%%%%%%%%%%%%%%%%%%%%%%%%%%%%%%%%%%%%%%%%%%%%%%%%%%%%%%%%%%%%%%%%%%
%%%%%%%%%%%%%%%%%%%%%%%%%%%%%%%%%%%%%%%%%%%%%%%%%%%%%%%%%%%%%%%%%%%%
%%%%%%%%%%%%%%%%%%%%%%%%%%%%%%%%%%%%%%%%%%%%%%%%%%%%%%%%%%%%%%%%%%%%
In this Appendix we derive the correction to the Cooperon self energy to
the fourth order in the exchange constant $J$. For this purpose we
will use the rules of the Keldysh diagrammatic technique, outlined in
Section \ref{sec:4.3}. 

The contribution to the Cooperon self energy is shown in
Fig.~\ref{fig:RPA}.  The electron-hole loops can be treated as the
self energy $\hat \varrho$ of the spin-spin Keldysh Green's function,
responsible for the spin relaxation as a result of the electron
inelastic scattering. The imaginary part of this electron-hole loop
determines the Korringa spin relaxation rate.\cite{Korringa} In the
Keldysh formalism, we can distinguish three components of the self
energy, related to the advanced, retarded and Keldysh elements of the
spin-spin Green's function.

We have for the difference of retarded and advanced components of the
spin-spin self energy the following expression
\begin{eqnarray}
\varrho^{\rm r}(\o)-\varrho^{\rm a}(\o) =2\pi\nu^2J^2\o,
\end{eqnarray}
and the Keldysh component is
\begin{eqnarray}
\varrho^{\rm k}(\o)=\coth\frac{\o}{2T}
\left(\varrho^{\rm r}(\o)-\varrho^{\rm a}(\o)\right).
\end{eqnarray}
This result is of no surprise. Indeed, the self energy is not related to
the type of interaction, since it is uniquely determined by the
electron-hole loop, rather than the interaction lines. 

Here we are interested in the strong magnetic field limit, $\os\gg T$,
so that only two impurity spin states with $S_z=S$ and $S_z=S-1$ are
involved. Under this condition, the diagrams in Fig.~\ref{fig:RPA} can
be represented in terms of the modified spin-spin correlators, which
to the second order in the exchange constant $\nu^2 J^2$ have the
form:
\begin{eqnarray}
{\cal L}_{1}^{\rm r,a}(\o) &= & {\cal L}_{+-}^{(0){\rm r,a}}(\o)
\varrho^{\rm r,a}(\o) {\cal L}_{+-}^{(0){\rm r,a}}(\o)
\\
{\cal L}_{1}^{\rm k}(\o) &= & {\cal L}_{+-}^{(0){\rm r}}(\o) \varrho^{\rm
k}(\o) {\cal L}_{+-}^{(0){\rm a}}(\o),
\end{eqnarray}
where ${\cal L}_{1}^{(0){\rm r,a}}(\o)=\langle S\rangle /(\o-\os\pm
i0)$ are the bare spin-spin correlation functions, defined by
Eq.~(\ref{scorr_pm}). Using this notations, we reduce diagrams in
Fig.~\ref{fig:RPA} to diagrams in Figs.~\ref{fig:Cse}a and
\ref{fig:Cse}b. Then, after the standard calculations, we obtain
\begin{eqnarray}
\Gamma_{4}(\e)=2\pi n_{\rm s} \nu^3 J^4  S^2\frac{\pi^2T^2+\e^2}{\os^2}.
\label{eq:Gamma4}
\end{eqnarray}
Using the definition of $\ts$, we rewrite Eq.~(\ref{eq:Gamma4}) in the
form of Eq.~(\ref{eq:SE2}).

%%%%%%%%%%%%%%%%%%%%%%%%%%%%%%%%%%%%%%%%%%%%%%%%%%%%%%%%%%%%%%%%%%%%%%%%
%%%%%%%%%%%%%%%%%%%%%%%%%%%%%%%%%%%%%%%%%%%%%%%%%%%%%%%%%%%%%%%%%%%%%%%%
\section{Conductance correlation function}
\label{app:D}
%%%%%%%%%%%%%%%%%%%%%%%%%%%%%%%%%%%%%%%%%%%%%%%%%%%%%%%%%%%%%%%%%%%%%%%%
%%%%%%%%%%%%%%%%%%%%%%%%%%%%%%%%%%%%%%%%%%%%%%%%%%%%%%%%%%%%%%%%%%%%%%%%

In this Appendix we outline the calculation of the conductance
correlation function ${\cal K}(\Df)$, see Eq.~(\ref{eq:CF-Hcs}). The
contributions to ${\cal K}(\Df)$ originate from two diagrams, shown in
Fig.~\ref{fig:CF-diag}. We have
\begin{eqnarray}
{\cal K}(\Df) & = &  
{\cal K}^{{\rm (a)}}(\Df) + {\cal K}^{{\rm (b)}}(\Df),
\label{eq:31}
\end{eqnarray}
where the first term ${\cal K}^{\rm (a)}(\Df)$ is the contribution
from fluctuations of the diffusion coefficient
\begin{eqnarray}
{\cal K}^{(\rm a)}(\Df)\! &\! =\! &\!
\alpha \frac{e^4}{(2\pi\hbar)^2} \frac{D^2}{R^4}\!\! 
\int\frac{d\e d\e'}{16T^2} \frac{1}{\cosh^2\!\!\e/2T}\frac{1}{\cosh^2\!\!\e'/2T}
\nonumber
\\
& \times &\!\!
2 \sum_{n=-\infty}^{+\infty}\!\!\!\sum_{i}\!\left| 
{\cal D}_i\left(\e,\e',q_n-\frac{\Df}{R\Phi_0}\right) \right|^2\!\!,
\label{eq:32a}
\end{eqnarray}
and second term in Eq.~(\ref{eq:31}) originates from fluctuations of
the electron density of states:
\begin{eqnarray}
{\cal K}^{(\rm b)}(\Df)\! &\! =\! &\!
\alpha \frac{e^4}{(2\pi\hbar)^2} \frac{D^2}{R^4}\!\! 
\int\frac{d\e d\e'}{16T^2} \frac{1}{\cosh^2\!\!\e/2T}\frac{1}{\cosh^2\!\!\e'/2T}
\nonumber
\\
& \times \!\!&\!\!\sum_{n=-\infty}^{+\infty}
\!\!\!\sum_{i} \! {\rm Re }\!\!\left\{ 
{\cal D}^2_i\!\!\left(\e,\e',q_n-\frac{\Df}{R\Phi_0}\right) \!\right\}\!\!.
\label{eq:34a}
\end{eqnarray}
Here ${\cal D}_i(\e,\e',q)$ is the $i$th diffuson component and the
sum over $n$ runs through all discreet values of momentum $q_n=n/R$.
We apply the Poisson summation formula
$$
\sum_{n=-\infty}^{+\infty}f(n)=\sum_{k=-\infty}^{\infty}\int
f(n)e^{2\pi i k n} dn
$$
to Eqs.~(\ref{eq:32a}) and (\ref{eq:34a}) and obtain
Eq.~(\ref{eq:CF-Hcs}) with amplitudes ${\cal K}_k$ defined by
Eqs.~(\ref{eq:31:D}) and (\ref{eq:MCF-R}).

\section{Diffuson decay rate due to scattering off quantum magnetic impurities}
\label{app:E}

We consider the singlet contribution to the conductance fluctuations
in the limit of short spin relaxation time $\tau_{\rm T}$, so that the
components of spin perpendicular to magnetic field are not correlated
at the characteristic time $\Dt$ between conductance measurements,
$f_\bot(\Dt)\to 0$ in Eqs.~(\ref{eq:500}). In this case the diffuson
self-energy is defined by diagrams similar to ones shown in
Fig.~(\ref{fig:Cse}), but with the opposite direction of the advanced
Green's function.

The diffuson self energy consists of three terms; two of them coincide
with the self energies $\Sigma^{\rm r}(\e)$ and $\Sigma^{\rm a}(\e)$
of retarded and advanced electron Green functions, respectively, and
the third term is the vertex correction $\Sigma^{\rm v}(\e)$.  Using
the formalism, presented in Appendix \ref{app:B}, we obtain the
following expression for the self energy of electron Green functions,
see Eqs.~(\ref{eq:Ese}):
\begin{eqnarray}
\Sigma^{\rm r,a}(\e) & = & 
\frac{1}{2\ts}\left[
1 -\frac{\langle \hat{S}_{z}\rangle^{2}
+\langle \hat{S}_{z}\rangle\tanh(\e+\omega_{\rm{s}})/2T}
{S(S+1)}
\right.
\nonumber
\\
& \pm &
\left.
    i\frac{\langle \hat{S}_{z}\rangle}{S(S+1)} \int
    \frac{\tanh(\e+\omega)/2T}
    {\omega+\omega_{\rm s}} \frac{d\omega}{2\pi}
\right] .
\label{eq:e.1}
\end{eqnarray}
The vertex correction to the diffuson originates from the impurity spin
correlator, taken at time difference $\Dt$, see
Eq.~(\ref{eq:30}). Since we assume that the impurity spin relaxation time is
much smaller than the delay time $\Dt$, the spin correlator is
$\chi_z=\Sodin ^2$ and we have
\begin{equation}
\Sigma^{\rm v}(\e)=\frac{1}{\ts}\Sodin^2.
\label{eq:e.2}
\end{equation}
 
Combining the electron self energy part and the vertex correction
given by Eqs.~(\ref{eq:e.1}) and (\ref{eq:e.2}), we obtain the
diffuson self energy in the form:
\begin{widetext}
\begin{eqnarray}
\Gamma(\e,\e') & = &
\frac{1}{\ts}\left[
1 -\frac{\langle \hat{S}_{z}\rangle^{2}}{S(S+1)}
-
\frac{\langle \hat{S}_{z}\rangle}{2S(S+1)}
\left(
    \tanh\frac{\e+\omega_{\rm{s}}}{2T}
    +\tanh\frac{\e'+\omega_{\rm{s}}}{2T}
\right)
\right.
\nonumber
\\
& + &
\left.
    i\langle \hat{S}_{z}\rangle\int
    \frac{\tanh(\e+\omega)/2T-\tanh(\e'+\omega)/2T}
    {\omega+\omega_{\rm s}} \frac{d\omega}{2\pi}
\right] .
\label{eq:dse}
\end{eqnarray}
For the calculation of the conductance fluctuations, we have to
consider a finite difference between the energies $\e$ and $\e'$,
transferred along the advanced and retarded Green's functions.  The
last term in Eq.~(\ref{eq:dse}) originates from the real part of the
electron self energy and vanishes if $\e=\e'$. This term represent
renormalization of the electron density of states.

The amplitude of the $k$th harmonic of the conductance correlation
function is expressed in terms of ${\cal A}^{\rm (4)}_k$ and ${\cal
  B}^{\rm (4)}_k$, see Eqs.~(\ref{eq:31:D}) and (\ref{eq:MCF-R}):
\begin{subequations}
\begin{eqnarray}
{\cal A}_k^{\rm (4)} & = & 
\sqrt{\frac{D}{R^2}}\frac{2\pi}{\e-\e'+{\rm Im}\Gamma(\e,\e')}
{\rm Im}\left\{
\frac{\exp\left(
-2\pi k R\sqrt{\gd+\Gamma(\e,\e')+i(\e-\e')}/\sqrt{D}
\right)}
{\sqrt{\gd+\Gamma(\e,\e')+i(\e-\e')}}
\right\}
\\
{\cal B}_k^{\rm (4)} & = &  
\sqrt{\frac{D\pi^3}{8R^2}}
{\rm Re}\left\{
\left(1+
2\pi k \frac{R\sqrt{D}}{\sqrt{\gd+\Gamma(\e,\e')+i(\e-\e')}}
\right)
\exp\left(
-2\pi k R\frac{\sqrt{\gd+\Gamma(\e,\e')+i(\e-\e')}}{\sqrt{D}}
\right)
\right\}.
\end{eqnarray} 
\end{subequations}
\end{widetext}

At high temperature $T\ts\gg 1$ integral in Eq.~(\ref{eq:31:D}) over
$\e-\e'$ converges at $|\e-\e'|\lesssim 1/\ts$, and we obtain
Eqs.~(\ref{11'}) and (\ref{eq:dqdt}).

%\end{multicols}
\end{document}